\documentclass[12pt]{article}
\pdfoutput=1

\usepackage{pgfplots}
\usetikzlibrary{decorations.pathmorphing}

\tikzset{snake it/.style={decorate, decoration=snake}}
\pgfplotsset{compat=1.10}
\usepgfplotslibrary{fillbetween}
\usetikzlibrary{patterns}

\usepackage{draft} 
\usepackage{hyperref}
\usepackage{graphicx,color,subfig}
\usepackage{cite}
\usepackage{mciteplus}
\usepackage{skak}
\usepackage{xcolor}
\usepackage{empheq}
\usepackage{tikz}
\DeclareFontFamily{OT1}{pzc}{}
\DeclareFontShape{OT1}{pzc}{m}{it}{<-> s * [1.10] pzcmi7t}{}
\DeclareMathAlphabet{\mathpzc}{OT1}{pzc}{m}{it}

\def\be#1\ee{\begin{align}#1\end{align}}



\newcommand{\IE}[0]{{\textit{i.e. }}}

\begin{document}

\unitlength = .8mm

\begin{titlepage}

\begin{center}

\hfill \\
\hfill \\
\vskip 1cm

\title{\Huge  Winding Tachyons and Stringy Black Holes}

\author{Bruno Balthazar, Jinwei Chu,
David Kutasov}
\address{
Kadanoff Center for Theoretical Physics and Enrico Fermi Institute\\ University of Chicago, Chicago IL 60637
}
\vskip 1cm

\email{brunobalthazar@uchicago.edu, jinweichu@uchicago.edu, dkutasov@uchicago.edu}

\end{center}

\abstract{We study string theory on $\mathbb{R}^d\times \mathbb{S}^1$. For applications to thermodynamics, the circumference of the $\mathbb{S}^1$ is the inverse temperature, $\beta$. We show that for $d=6$, the low energy effective field theory at the inverse Hagedorn temperature, $\beta=\beta_H$,  has a one parameter family of normalizable spherically symmetric solutions that break the winding symmetry around the $\mathbb{S}^1$. The resulting backgrounds exhibit an enhanced symmetry, with the symmetry breaking pattern $SU(2)_L\times SU(2)_R\to SU(2)_{\rm diagonal}$. The effective field theory analysis of these backgrounds is reliable for some range of parameters. More generally, they are described by a  worldsheet CFT, which corresponds to the free theory on $\mathbb{R}^6\times \mathbb{S}^1$ perturbed by a non-abelian Thirring deformation with an $r$-dependent coupling. We propose that, in a certain scaling limit, string theory in these backgrounds is described by the $SL(2,\mathbb{R})/U(1)$ cigar, and provides a thermodynamic description of weakly coupled highly excited fundamental strings. We also discuss the relation of these backgrounds to Euclidean black holes with near-Hagedorn Hawking temperature, and possible generalizations to other $d$.
}

\vfill

\end{titlepage}

\eject

\begingroup
\hypersetup{linkcolor=black}
\tableofcontents
\endgroup

\section{Introduction}
\label{sec:intro}
Large Schwarzschild black holes are believed to describe generic high energy states in quantum gravity. Their Euclidean version plays an important role in the thermodynamics of the theory. The Euclidean solution is better behaved since it does not have a curvature singularity; we will focus here on this case. 

In the context of string theory, for large Schwarzschild radius $R_{sch}\gg l_s$, we can view the Euclidean Schwarzschild black hole as a solution of Einstein gravity, but it is more properly thought of as a worldsheet Conformal Field Theory (CFT). As the Schwarzschild radius decreases, the size of stringy corrections to the GR picture increases~\cite{Callan:1988hs,Myers:1987qx,Mertens:2015ola,Chen:2021qrz,Moura:2009it,Moura:2018psq}, and for $R_{sch}\sim l_s$  one needs to use the tools of string theory to describe it.

The Schwarzschild CFT  breaks the $U(1)_L\times U(1)_R$ symmetry associated with momentum and winding around the Euclidean time circle down to the diagonal $U(1)$ associated with momentum -- i.e. it breaks the winding symmetry. Indeed, a string wound around the Euclidean time circle can unwind at the Euclidean horizon. In the CFT  there is another important winding symmetry breaking effect, that is not visible in the gravity approximation -- the condensate of the closed string tachyon winding around the Euclidean time circle; see e.g. ~\cite{Kutasov:2005rr,Chen:2021dsw}. At large $R_{sch}$, this condensate can be viewed as a non-perturbative $\alpha'$ effect which provides a small correction to the GR picture, but for string size black holes it is significant. 

The tachyon winding once around the Euclidean time circle becomes massless at the Hagedorn temperature of the flat spacetime string theory~\cite{Atick:1988si}. Thus, to study Euclidean Black Holes (EBH's) in that regime using a low energy Effective Field Theory (EFT), we need to include it in the effective Lagrangian. The expectation value of the tachyon goes to zero at large radial distance, $r$, and the geometry becomes flat there, so the EFT analysis should be reliable far from the Euclidean horizon. When we approach the horizon, in general we expect the EFT to break down, because both the curvature and the tachyon condensate are large there. 

Horowitz and Polchinski\footnote{The analysis of HP was recently generalized from $\mathbb{R}^d\times \mathbb{S}^1$ to thermal $AdS_{d+1}$ in \cite{Urbach:2022xzw}.} (HP) found \cite{Horowitz:1997jc} that for $d<6$ non-compact space dimensions, the above low energy EFT has a spherically symmetric solution that breaks the winding symmetry via a condensate of the winding tachyon, and remains controlled all the way to small $r$. This solution exists when the temperature $T$ is slightly below the Hagedorn temperature $T_H$, and goes to zero as $T\to T_H$. The small parameter that controls the validity of the EFT is $(T_H-T)/T_H$. In section \ref{sec:hpsol} we review this solution and its thermodynamic properties. 

The Hagedorn temperature of string theory, $T_H=1/\beta_H$, controls the high energy density of perturbative string states,
\begin{equation}
\label{hagent}
	\begin{split}
    n(E)=e^{S(E)}\sim \frac{e^{\beta_H E}}{E^{1+\frac d2}}~.
	\end{split}
	\end{equation}
A natural question is whether the large entropy $S(E)$ \eqref{hagent} can be described by a Euclidean solution of string theory, such as the HP solution, or the continuation of the EBH to $T\simeq T_H$, via an analog of black hole thermodynamics. In this paper we discuss this issue, using the EFT as a guide. 

In section \ref{sec:d6}, we describe a solution of the EFT at $T=T_H$ for $d=6$, that is labeled by the value of the winding tachyon condensate at the origin, $\chi(0)$. This parameter can take any (real positive) value, due to a scaling symmetry of the EFT. From the thermodynamic perspective, $\chi(0)$ can be thought of as determining the mass of the solution, $M$, the analog of the black hole mass in black hole thermodynamics. The fact that the temperature is independent of the mass for this solution implies that it describes a system with Hagedorn entropy.

The EFT analysis of this solution is reliable for small $\chi(0)$. This region corresponds to large mass, $ M\gg M_{\mathrm{corr}}\equiv\frac{m_s}{g_s^2}$ ($M_{\mathrm{corr}}$ is the string-black hole correspondence mass \cite{Horowitz:1996nw}). As $\chi(0)$ increases, i.e. $M$ decreases, the corrections to the EFT grow, and eventually, when $M$ approaches $M_{\rm corr}$, the EFT analysis breaks down. In that region, one needs to analyze the solution using the tools of string theory.  

In section \ref{sec:ws}, we describe the worldsheet theory that corresponds to the solution of section \ref{sec:d6} for general $\chi(0)$. We show that it is given by a non-abelian Thirring model with a coupling that depends on $r$. The $r$ dependence of the coupling is determined by the requirement that the theory is conformal. The resulting background can be thought of as describing a localized spherical bubble of non-abelian Thirring -- the Thirring coupling approaches a finite value at $r=0$, and goes to zero at large $r$. 

One of the interesting features of this system is that it exhibits an enhanced symmetry. The usual pattern of symmetry breaking, $U(1)_L\times U(1)_R\to U(1)_{\rm diagonal}$, associated with winding violating backgrounds like the EBH and HP solutions, is replaced in this case by $SU(2)_L\times SU(2)_R\to SU(2)_{\rm diagonal}$. We expect this to be a general feature of winding violating backgrounds at the Hagedorn temperature. 

In section \ref{sec:md6} we discuss the relation of the solution of sections \ref{sec:d6}, \ref{sec:ws} to small Euclidean black holes. We argue that if we continue the mass $M$ of the solution to the region $M\ll M_{\rm corr}$, the background develops a long throat located near $r=0$, in which, after reducing on the sphere  $\mathbb{S}^5$, it is described by the Euclidean two dimensional black hole, $SL(2,\mathbb{R})/U(1)$ \cite{Witten:1991yr}. As evidence, we point out that the two dimensional black hole has the same symmetry breaking pattern as our solution. 

We also argue that continuing the large Euclidean Schwarzschild black hole to small mass gives a solution that approaches that of sections \ref{sec:d6}, \ref{sec:ws}. Thus, in this limit, the EBH gives rise to a background whose thermodynamics approaches that of highly excited weakly coupled fundamental strings. 

In section \ref{sec:discu} we discuss possible generalizations of the picture of the previous sections to $d\not=6$. For $d<6$ we take the attitude that the EBH and HP solutions should extend to solutions of the full theory, labeled by the temperature, even in regions where the relevant EFT that was used for their construction fails. We also assume that the behavior of the solutions depends smoothly on $d$.

The simplest deformation of the structure in $d=6$ is one where the EBH and HP solutions are connected at some finite $R-R_H$ and $M$ (see figure \ref{dneq6thermo}). This seems to disagree with the results of \cite{Chen:2021dsw}, that find that such a continuous transition is impossible in classical type II string theory. If this is the case, one needs to understand the behavior of the HP solution as the temperature decreases.\footnote{The authors of \cite{Chen:2021dsw} suggest that the transition could be via a singular CFT, but this seems unnatural since at the transition the mass $M$ and temperature $T$ are in the regime that should be well described by worldsheet CFT.}

We also briefly discuss the case $d>6$. We point out that a natural scenario is that the structure is the same as that of the $d=6$ case, but it is also possible that the $T=T_H$ solution ceases to exist for $d>6$. This would imply a discontinuity in going from $d=6$ to $d=6+\epsilon$ for arbitrarily small $\epsilon$. 

We include an appendix that contains some technical results used in the text.

\section{The Horowitz-Polchinski solution}
\label{sec:hpsol}
We start with a Euclidean spacetime of the form $\mathbb{R}^d\times \mathbb{S}^1$. The $\mathbb{S}^1$ is Euclidean time; its circumference is  
    \begin{equation}
	\begin{split}
    \beta=2\pi R=1/T\ ,
	\end{split}
	\end{equation}
the inverse temperature. We will study the theory for temperatures in the vicinity of the Hagedorn temperature $T_H$. The inverse Hagedorn temperature, $\beta_H=1/T_H$, can be written as~\cite{Atick:1988si}
\begin{equation}
	\label{tthh}
	\begin{split}
    \beta_H=2\pi R_H\ , \;\;R_H^{\rm bosonic}=2l_s\ ,\;\; R_H^{\rm type\; II}=\sqrt2l_s\ .
	\end{split}
	\end{equation}
When the temperature is slightly below $T_H$, \IE $R$ is slightly above $R_H$, the tachyon with winding one, viewed as a complex field $\chi(x)$ on $\mathbb{R}^d$, becomes light. Its mass is given by
\ie
	\label{mR}
    m^2_\infty=\frac{R^2-R^2_H}{\alpha'^2}\ ,
\fe
and it goes to zero as $R\to R_H$. We will be interested in the regime $m_\infty\ll m_s\equiv 1/\sqrt{\alpha'}$.

The EFT obtained by reducing on the Euclidean time circle is described by the action
\ie
\label{fac}
I_d=\frac{\beta}{16\pi G_N}\int d^dx\sqrt{g}e^{-2\phi_d}\left[-\mathcal{R}-4(\nabla \phi_d)^2+(\nabla\varphi)^2+|\nabla \chi|^2+m^2_\infty|\chi|^2+\cdots\right]\ .
\fe
Here $g$ is the metric on $\mathbb{R}^d$, $\phi_d$ is the $d$-dimensional dilaton, $G_N$ is the $d+1$ dimensional Newton's constant, and $\mathcal{R}$ is the scalar curvature of $g$. The scalar field $\varphi$ controls the radius of the Euclidean time circle. The ellipsis in \eqref{fac} stands for other fields, that will not be important for the analysis, and for higher order terms in the fields and in derivatives. 

As mentioned above, we are interested in solutions of the equations of motion that have a non-zero expectation value of the field $\chi$, and thus break the winding symmetry. For small $\chi$ one can neglect the back-reaction on the metric $g$ and dilaton $\phi_d$. Thus, we will set them to the standard flat space values. 

The field $\chi$ satisfies the Klein-Gordon equation 
\begin{equation}
	\label{chir}
	\begin{split}
    \nabla^2\chi=m^2_\infty\chi \ .
	\end{split}
	\end{equation}
We will focus on spherically symmetric solutions of \eqref{chir}, which behave at large $r$ like $\chi(r)\sim r^{-\frac{d-1}{2}}\exp(-m_\infty r)$. 

To study the HP solution, as well as near-Hagedorn black holes, we need to include an additional term in the effective Lagrangian \eqref{fac} -- the coupling of the radion field $\varphi$ to the winding tachyon $\chi$. This coupling owes its existence to the fact that the mass of $\chi$ depends on the radius of the thermal circle, which depends on $\varphi$. We will use conventions where positive $\varphi$ corresponds to a larger circle. The leading coupling is proportional to $\varphi|\chi|^2$.

Putting all these elements together, we arrive at the action
	\begin{equation}
	\label{Sphichi}
	\begin{split}
 	I_d=\frac{\beta}{16\pi G_N}\int d^dx\left[(\nabla\varphi)^2+|\nabla \chi|^2+(m_{\infty}^2+\frac{\kappa}{\alpha'}\varphi)|\chi|^2\right]\ .
	\end{split}
	\end{equation}
Here we used the conventions of \cite{Chen:2021dsw}. In particular, the constant $\kappa$ is given by 
\begin{equation}
	\label{kappabt}
	\begin{split}
    \kappa^{\rm bosonic}=8\ ,\;\; \kappa^{\rm type\; II}=4\ .
	\end{split}
	\end{equation}
We note for future reference that the fields $\varphi$ and $\chi$ are dimensionless, and their relative normalization is fixed by the requirement that their kinetic terms are equal.

The equations of motion of \eqref{Sphichi} are 
	\begin{equation}
	\label{eom0}
	\begin{split}
 	\nabla^2\chi &=(m_{\infty}^2+\frac{\kappa}{\alpha'}\varphi)\chi \ ,\\
 	\nabla^2\varphi &=\frac{\kappa}{2\alpha'}|\chi|^2\ .
	\end{split}
	\end{equation}
For $m_\infty=0$, these equations have a dilatation symmetry, that we will use extensively below. Under this symmetry, $\varphi$ and $\chi$ have scaling dimension two and $x$ has scaling dimension $-1$. The mass $m_\infty$ can be thought of as a coupling of scaling dimension one. In other words, it is a relevant coupling, whose effects become important at large $x$, and go to zero at small $x$ (relative to $1/m_\infty$). 

On the other hand, higher order contributions to the action \eqref{Sphichi}, such as $|\chi|^4$, $\varphi^2|\chi|^2$, etc, give rise to irrelevant couplings, that are unimportant at large $x$. Conversely, such terms play an important role at small $x$, and the action \eqref{Sphichi} is in general expected to break down there. 

Note that the above dilatation symmetry is in general distinct from the usual symmetry of free massless Klein-Gordon theory, under which scalar fields such as $\chi$ and $\varphi$ have scaling dimension $(d-2)/2$. This is related to the fact that the action \eqref{Sphichi} is not invariant under it, but rather is rescaled by a $d$-dependent overall factor. For $d=6$ this rescaling factor disappears, and the two symmetries coincide. This is probably related to the special role that $d=6$ plays in the HP analysis, as well as in our discussion below, though the precise relation is unclear to us.

The Horowitz-Polchinski solution \cite{Horowitz:1997jc} is a spherically symmetric solution of \eqref{eom0}, for which $\varphi(r)$,  $\chi(r)$ go to zero at infinity and approach constant values at the origin. The phase of the complex field $\chi$ is fixed in this solution, and we can take $\chi(r)$ to be real and positive, without loss of generality. 
\begin{figure}[h!]
\centering
 {\subfloat{\includegraphics[width=0.48\textwidth]{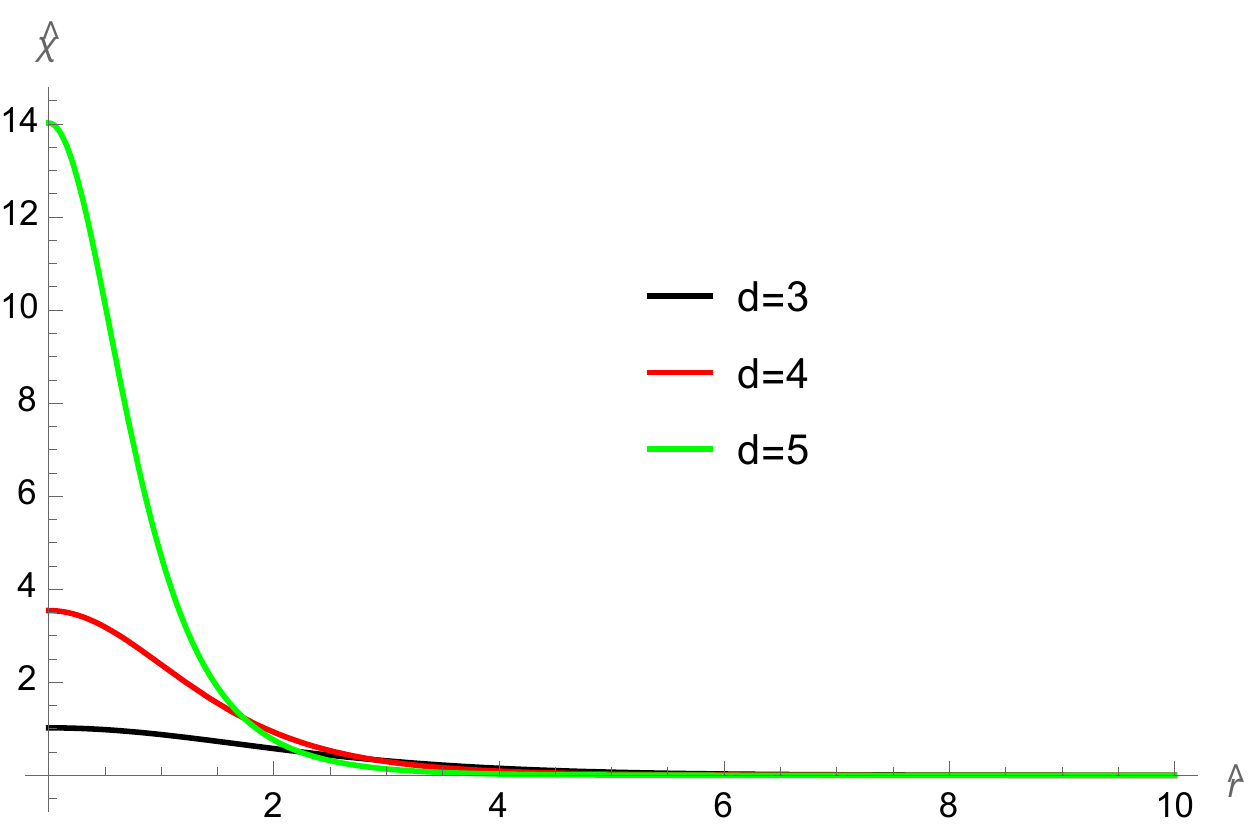}\label{chifig}}}~ 
 {\subfloat{\includegraphics[width=0.48\textwidth]{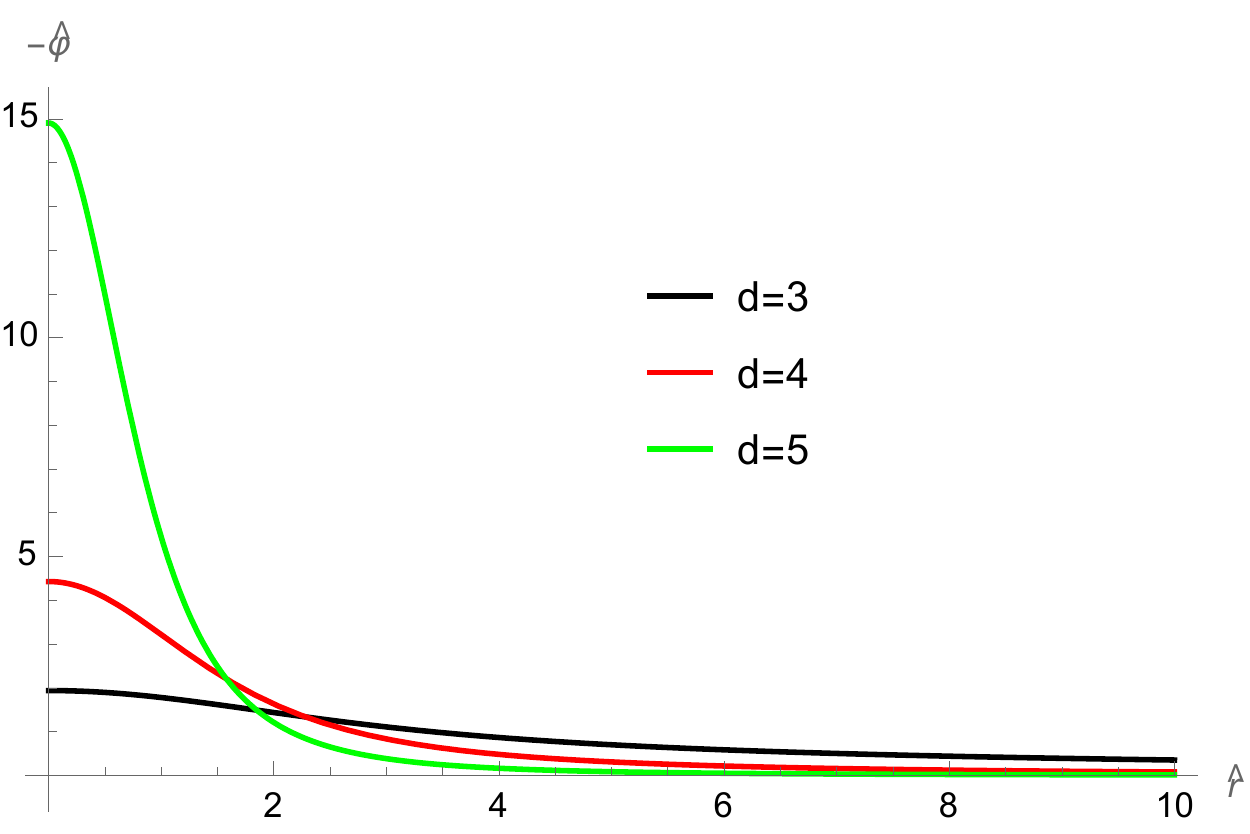}\label{phifig}}} 
 \caption{The profiles of $\hat\chi$ and $-\hat\varphi$ (defined in \eqref{rescaled}) for $d=3,4,5$.}
 \label{chiphifigs}
\end{figure}

The solution $\chi(r)$ is then a monotonically decreasing function of $r$. The radion $\varphi(r)$ is negative and monotonically increasing -- the radius of the thermal circle decreases as $\chi$ increases, due to the effect of the condensate of the winding string. It can be expressed in terms of $\chi$ as 
\begin{equation}
	\label{phichi}
	\begin{split}
    \varphi(x)=-\frac{\kappa}{2(d-2)\omega_{d-1}\alpha'}\int d^dy\frac{\chi^2(y)}{|x-y|^{d-2}}\ ,
	\end{split}
	\end{equation}
where $\omega_{d-1}$ is the area of the unit $(d-1)$-sphere.

\begin{figure}[h!]
\centering
 {\subfloat{\includegraphics[width=0.48\textwidth]{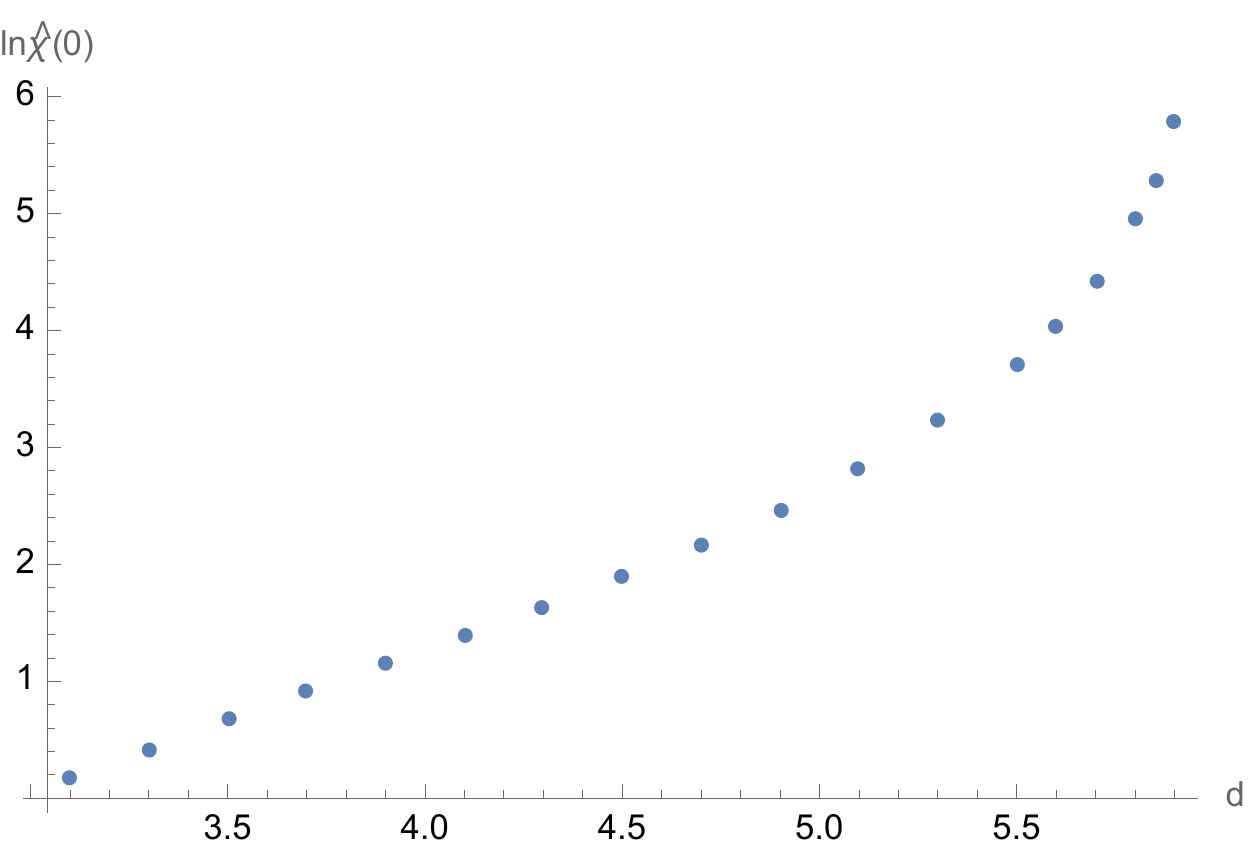}\label{chi0fig}}}~ 
 {\subfloat{\includegraphics[width=0.48\textwidth]{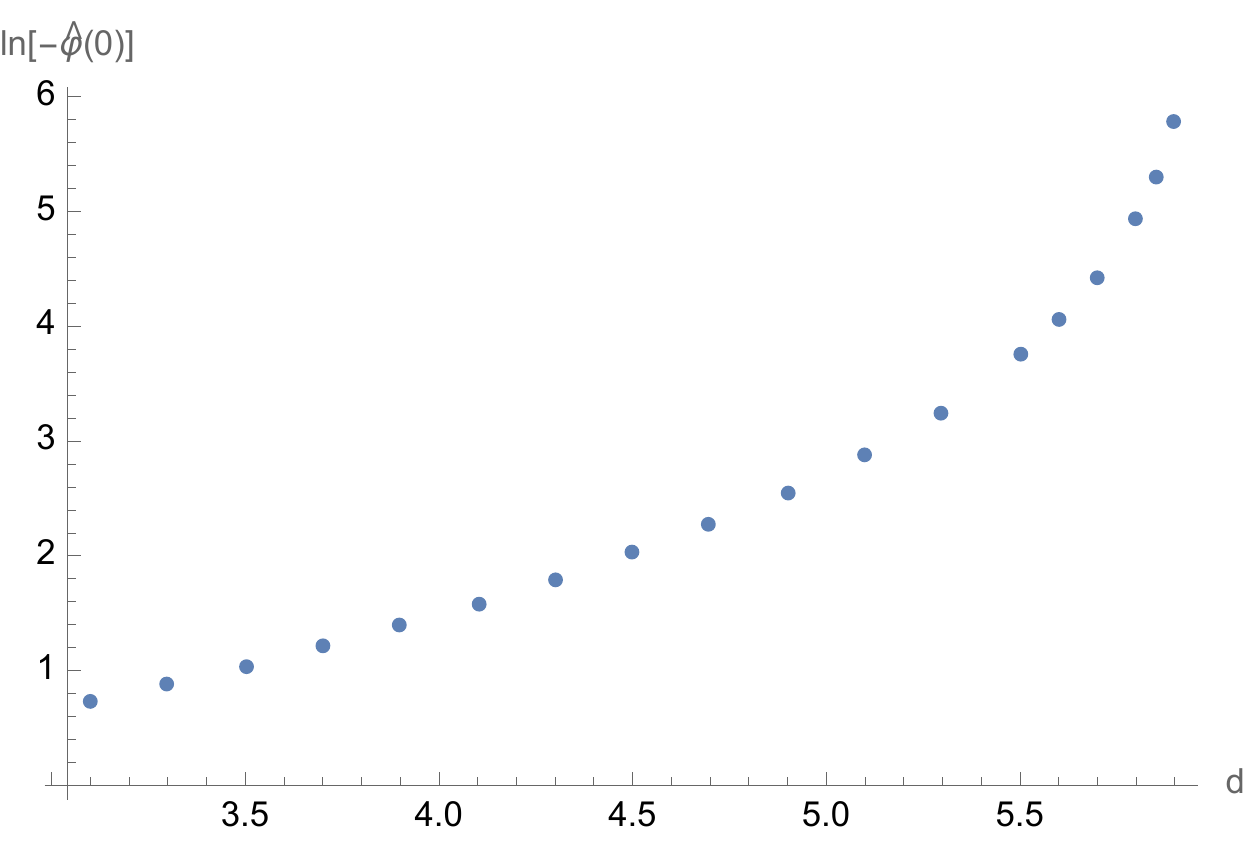}\label{phi0fig}}} 
 \caption{$\hat{\chi}(0)$ and $-\hat{\varphi}(0)$ as a function of $d$.}
 \label{chiphi0figs}
\end{figure}

To solve equations \eqref{eom0}, we use the scaling symmetry to define
\ie
	\label{rescaled}
	x&=\hat{x}/m_\infty\ ,\\
 	\chi (x)&=\frac{\sqrt{2}\alpha' }{\kappa}m^2_{\infty}\hat{\chi}(\hat{x})\ ,\\
 	\varphi (x)&=\frac{\alpha'}{\kappa }m^2_{\infty}\hat{\varphi}(\hat{x})\ .
\fe
In terms of the hatted variables, these equations take the form (for spherically symmetric configurations $\hat\chi=\hat\chi(\hat r)$, $\hat\varphi=\hat\varphi(\hat r)$)
\begin{equation}
	\label{de}
	\begin{split}
 	\hat\nabla^2\hat\chi=\hat{\chi}''+\frac{d-1}{\hat{r}}\hat{\chi}' &=(1+\hat{\varphi})\hat{\chi} \ ,\\
 	\hat\nabla^2\hat\varphi=\hat{\varphi}''+\frac{d-1}{\hat{r}}\hat{\varphi}' &=\hat{\chi}^2\ .
	\end{split}
	\end{equation}
Equations \eqref{de} have a unique normalizable solution with the properties listed above for fixed $d$ (see figure \ref{chiphifigs}). It is instructive to plot $\hat{\chi}(0)$, $\hat{\varphi}(0)$ as a function of $d$, viewed as a continuous variable. We exhibit the result in figure \ref{chiphi0figs}. It has an interesting behavior near $d=6$, which we will discuss later. 

Equation \eqref{rescaled} implies that $\chi(0)$, $\varphi(0)$ scale like $m_\infty^2$. Therefore, for small $m_\infty$ the higher order corrections to the effective action \eqref{Sphichi} mentioned above can be neglected. When $m_\infty$ grows, such terms need to be included, and this EFT breaks down. This happens when $m_\infty\sim m_s$; thus the HP analysis is only valid for temperatures $T_H-T\ll m_s$.  

A useful way of parametrizing the solution of \eqref{eom0} is in terms of the behavior of $\varphi(r)$ at large $r$. Since $\chi(r)$ goes exponentially to zero at large $r$, the second equation in \eqref{eom0} implies that 
\begin{equation}
	\label{phias}
	\begin{split}
    \varphi(r)\sim -\frac{C_\varphi}{r^{d-2}}
	\end{split}
	\end{equation}
at large $r$. Taking $|x|$ to be large in (\ref{phichi}), we learn that
\begin{equation}
\label{Cphi}
    \begin{split}
    C_\varphi=\frac{\kappa}{2(d-2)\omega_{d-1}\alpha'}\int d^dx\ \chi^2(x)\ .    
    \end{split}
\end{equation}
Using \eqref{rescaled}, or equivalently the dilatation symmetry mentioned after equation \eqref{eom0}, we find that the constant $C_\varphi$ scales like
\begin{equation}
	\label{scaleCphi}
	\begin{split}
    C_\varphi\sim \alpha'\left(m_\infty\right)^{4-d}~.
	\end{split}
	\end{equation}
As we review below, the parametrization \eqref{phias} is useful since $C_\varphi$ is proportional to the energy of the solution. 

The asymptotic behavior of the field $\chi(r)$ can be deduced from the first equation in \eqref{eom0}. For $d>3$, it is given by 
\begin{equation}
	\label{chias}
	\begin{split}
    \chi(r)\sim A_\chi r^{-\frac{d-1}{2}}e^{-m_\infty r}\ .
	\end{split}
	\end{equation}
As before, we can determine the dependence of $A_\chi$ on $m_\infty$ from scaling,
\begin{equation}
	\label{scaleCchi}
	\begin{split}
    A_\chi\sim\alpha' \left(m_\infty\right)^{\frac{5-d}{2}}\ .
	\end{split}
	\end{equation}

We now review the thermodynamic properties of the Horowitz-Polchinski solution \cite{Horowitz:1997jc,Chen:2021dsw}. The ADM energy of the solution is given by
\begin{equation}
\label{massHP}
\begin{split}
M_{\mathrm{HP}}&=2(d-2)\omega_{d-1}\frac{C_\varphi}{16\pi G_N}\ .
\end{split}
\end{equation}
As mentioned above, the energy of the HP solution is proportional to the constant $C_\varphi$, which is determined by the temperature via equations \eqref{mR} and \eqref{scaleCphi}.

Another interesting feature of these solutions is that their entropy is non-zero, and of order $\frac{1}{G_N}$. It is given by
\begin{equation}
\label{entropyHP}
\begin{split}
S_{\mathrm{HP}}&=\frac{\kappa}{\alpha'}\frac{\beta_H}{16\pi G_N}\int d^d x \chi^2(x)
\\
&=2(d-2)\omega_{d-1}\frac{\beta_H C_\varphi}{16\pi G_N}\ .
\end{split}
\end{equation}
To leading order in $m_\infty$, it is related to the energy \eqref{massHP} via the relation $S_{\mathrm{HP}}=\beta_H M_{\mathrm{HP}}$, since $T\to T_H$ as $m_\infty\to 0$. Using the thermodynamic relation
\begin{equation}
    \begin{split}
        \frac{dS}{dM}=\beta=\beta_H+(\beta-\beta_H)\;,
    \end{split}
\end{equation}
we can obtain the leading correction away from Hagedorn behaviour by expressing $\beta-\beta_H$ in terms of $M_{\mathrm{HP}}$ using \eqref{scaleCphi}, \eqref{massHP}, and integrating in $M_{\mathrm{HP}}$. We find
\begin{equation}
\label{SHPcorr}
    \begin{split}
        S_{\mathrm{HP}}\left(M_{\mathrm{HP}}\right)=\beta_H M_{\mathrm{HP}}+\xi_d M_{\mathrm{HP}}^{\frac{6-d}{4-d}},~~~~~\xi_d=\frac{4-d}{6-d}\frac{(2\pi\alpha')^2}{2\beta_H}\left[\frac{16\pi G_N}{2(d-2)\omega_{d-1}}\frac{m^{4-d}_\infty}{C_\varphi}\right]^{\frac{2}{4-d}}.
    \end{split}
\end{equation}
Note that by \eqref{scaleCphi}, $\xi_d$ is independent of $\beta$, but it does depend on the overall coefficient in that equation. This coefficient can be determined using the numerical solution. At $d=4$, $M_{\mathrm{HP}}$ is independent of $\beta$, and therefore the argument used above fails, so we cannot use it to obtain the leading correction to the entropy-energy relation.

The free energy is given by evaluating the action \eqref{Sphichi} on the Horowitz-Polchinski solution, 
\begin{equation}
\label{FHP}
    \begin{split}
         F=\frac{I_d}{\beta}&=\frac{2}{6-d}\frac{1}{16\pi G_N}m^2_\infty\int d^d x \chi^2(x)
        \\
        &=\frac{4}{\kappa}\omega_{d-1}\frac{d-2}{6-d}\alpha' m^2_\infty\frac{C_\varphi}{16\pi G_N}
        \\
        &\sim \alpha'^2 m^{6-d}_\infty\frac{1}{G_N}~,
    \end{split}
\end{equation}
where on the last line we used \eqref{scaleCphi}. The free energy can also be obtained from the thermodynamic relation $\beta F=\beta M-S$, using \eqref{massHP} and \eqref{SHPcorr} for $M$ and $S$ respectively.

So far, we discussed the regime of validity of the EFT \eqref{Sphichi} in classical string theory. It is also interesting to ask when quantum $(g_s)$ corrections become significant. The action $I_d$ \eqref{FHP} scales like $\left(m_\infty\right)^{6-d}$. Thus, the classical EFT analysis is reliable when
\begin{equation}
	\label{gscor}
	\begin{split}
    g_s^\frac{2}{6-d}\ll \frac{m_\infty}{m_s}\ll 1~.
	\end{split}
	\end{equation}

For $4<d<6$, the mass of the HP solution, $M_{\mathrm{HP}} $ \eqref{massHP}, where $C_\varphi$ is given by \eqref{scaleCphi}, has the property that in the regime \eqref{gscor}, $M_{\mathrm{HP}}\gg M_{\mathrm{corr}}$. Here, $M_{\mathrm{corr}}$ is the Horowitz-Polchinski string/black-hole correspondence mass \cite{Horowitz:1996nw}, 
\begin{equation}
\label{mcorr}
    \begin{split}
        M_{\rm corr}=\frac{m_s}{g_s^2}~.
    \end{split}
\end{equation}
Thus, the HP solution is rather different from Euclidean Schwarzschild as a solution of the EFT \eqref{Sphichi}, since the latter has a much lower temperature for masses $M\gg M_{\mathrm{corr}}$. In particular, the former can be studied reliably using this EFT, while the latter is outside its regime of validity.  
We will comment later on the possibility that the two solutions are continuously connected at $(T_H-T)/T_H$ of order one, which is outside of the regime of validity of the HP Lagrangian \eqref{Sphichi} (and of the GR description of the EBH). See \cite{Chen:2021dsw} for a recent discussion.
	
For $d<4$, the mass \eqref{massHP} decreases as $T$ approaches $T_H$, and again it is possible that the EBH is continuously connected to the HP solution, with the transition between the two occuring at $(T_H-T)/T_H$ of order one. For $d>6$, the HP solution does not exist, while the EBH is expected to exist at temperatures close to the Hagedorn one. It corresponds to some solution of \eqref{eom0} far from the Euclidean horizon, which leaves the regime of validity of these equations near the horizon.

\section{$T=T_H$ for $d=6$}
\label{sec:d6}

As discussed in \cite{Horowitz:1997jc,Chen:2021dsw}, the HP solution does not exist for $d\ge 6$ and $m_\infty>0$. In this section we will see that a solution does exist for $d=6$ and $m_\infty=0$. We will also show that this solution can be obtained by taking a double-scaling limit of the HP solution.  

\subsection{The $d=6$ solution}
\label{sec:classical sols}

For $m_\infty=0$, the equations of motion \eqref{eom0} take the form 
\begin{equation}
	\label{eom6}
	\begin{split}
 	\nabla^2\chi &=\frac{\kappa}{\alpha'}\varphi\chi \ ,\\
 	\nabla^2\varphi &=\frac{\kappa}{2\alpha'}|\chi|^2\ .
	\end{split}
	\end{equation}
As mentioned in the previous section, these equations admit a scaling symmetry under which $\chi,\varphi$ have dimension 2. When $d=6$, this is also the (classical) dimension of the fields in the action \eqref{Sphichi}. In fact, in that case this scaling symmetry is enhanced to six dimensional conformal symmetry. 

Following the discussion of the previous section, we are looking for a normalizable solution $(\chi(r),\varphi(r))$, with a qualitative structure similar to that of figure \ref{chiphifigs}. In particular, we demand that $\chi(r),-\varphi(r)$ are positive and monotonic for all $r$, and that they go to positive constants as $r\to 0$, and to zero as $r\to\infty$. We next show that for such a solution 
\ie
\label{chieqphi}
\chi(r)+\sqrt2\varphi(r)=0
\fe
for all $r$. To see that, we note that a linear combination of the two equations in \eqref{eom6} is 
\ie
\label{lincomb}
\nabla^2(\chi+\sqrt2\varphi) =\frac{\kappa}{\sqrt2\alpha'}(\chi+\sqrt2\varphi)\chi~.
\fe
Multiplying this equation by $r^{d-1}$ and integrating from $0$ to $r_0$, we have\footnote{In writing \eqref{integcond}, we dropped the boundary term $\left.\left(\chi'(r)+\sqrt2\varphi'(r)\right)r^{d-1}\right|_{r=0}$, which vanishes for normalizable fields $\chi(r),\varphi(r)$ satisfying \eqref{eom6}.}
\ie
\label{integcond}
\left(\chi'(r_0)+\sqrt2\varphi'(r_0)\right)r_0^{d-1} =\frac{\kappa}{\sqrt2\alpha'}\int_0^{r_0} dr r^{d-1}(\chi+\sqrt2\varphi)\chi~.
\fe
We can now discuss separately the cases where $\chi(0)+\sqrt2\varphi(0)$ is positive, negative and zero. If it's positive, the integral on the r.h.s. of \eqref{integcond} is positive for small $r_0$, and therefore $\chi'(r_0)+\sqrt2\varphi'(r_0)$ is positive as well. Thus, $\chi+\sqrt2\varphi$ is an increasing function of $r$. In fact, this function cannot have a local maximum at any finite $r_0$, since at the first such maximum, the l.h.s. of \eqref{integcond} vanishes, while the r.h.s. is positive. 

This behavior is inconsistent with the fact that $\chi$ and $\varphi$ must go to zero at large $r$. Hence, $\chi(0)+\sqrt2\varphi(0)$ cannot be positive. Similarly, it cannot be negative. Thus, we conclude that $\chi(0)+\sqrt2\varphi(0)=0$. In fact, the above argument can be used to show that \eqref{chieqphi} is valid for all $r$, since otherwise this function would monotonically increase (in absolute value) as $r\to\infty$, and could not vanish there.

An alternative argument for \eqref{chieqphi} is to note that taking $\chi(r)$ as given, eq. \eqref{lincomb} can be thought of as a bound state equation in quantum mechanics for a zero energy s-wave state with wavefunction $\chi+\sqrt2\varphi$ in a spherically symmetric potential proportional to $\chi(r)$. Since $\chi(r)$ is assumed to be positive for all finite $r$, such a wavefunction must vanish. 

Plugging \eqref{chieqphi} into \eqref{eom6}, we see that these two equations reduce to
\ie
\label{neweom}
\nabla^2\chi =-\frac{\kappa}{\sqrt2\alpha'}\chi^2~.
\fe
This equation is invariant under the dilatation symmetry mentioned above. Thus, if $\chi(r)$ is a solution, so is $\lambda^2\chi(\lambda r)$ for all real, positive $\lambda$.  

Looking back at \eqref{phias}, we are seeking solutions that behave at large $r$ like 
\begin{equation}
	\label{dsixas}
	\begin{split}
    \chi(r)\sim \frac{C_\chi}{r^4}
	\end{split}
	\end{equation}
and go to a constant, $\chi(0)$, as $r\to0$. The value of $C_\chi$ can be changed by using the scaling symmetry mentioned above. This symmetry also relates $C_\chi$ and $\chi(0)$; it implies that $C_\chi\sim1/\chi(0)$. 

\begin{figure}[h!]
\centering
 \includegraphics[width=0.5\textwidth]{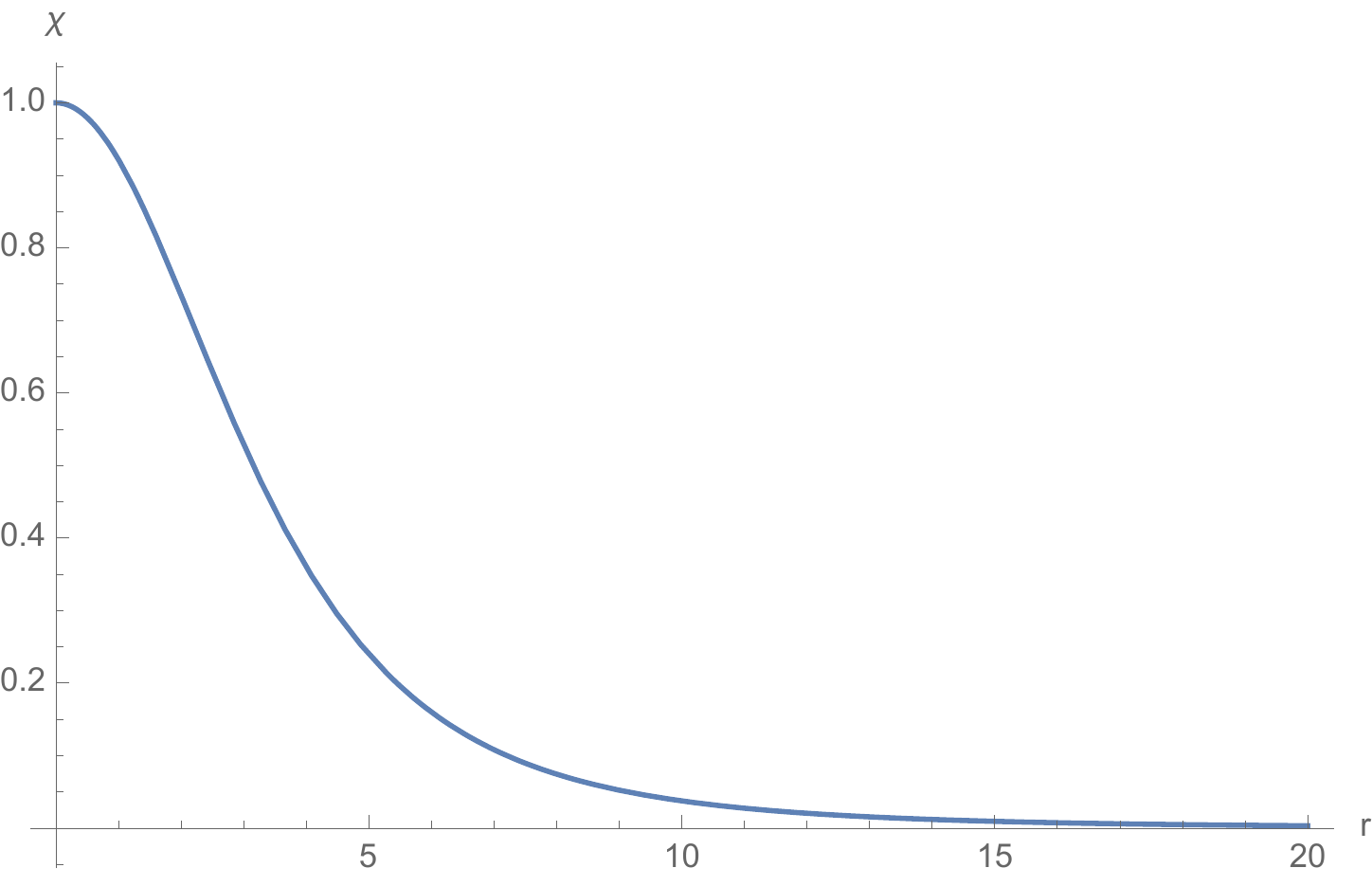}
 \caption{$\chi(r)$ profile for the solution \eqref{chisol}, for $\chi(0)=1$ and  $\frac{\sqrt{2}\alpha'}{\kappa}=1$.}
\label{d6solplot}
\end{figure}

The requisite solution of \eqref{neweom} is known analytically \cite{Mckane:1978me}. It is given by 
\begin{equation}
\label{chisol}
    \begin{split}
        \chi(r)=\frac{C_\chi}{\left(\frac{C_\chi\kappa}{24\sqrt{2}\alpha'}+r^2\right)^2}~.
    \end{split}
\end{equation}
As expected, it goes to a constant 
\begin{equation}
	\label{chizz}
	\begin{split}
    \chi(0)=\left(\frac{24\sqrt{2}\alpha'}{\kappa}\right)^2\frac{1}{C_\chi}
	\end{split}
	\end{equation}
as $r\to0$, and behaves like \eqref{dsixas} as $r\to\infty$, see figure \ref{d6solplot}.

Equation \eqref{chisol} describes a set of solutions at the Hagedorn temperature, $T=T_H$, labeled by the real positive constant $C_\chi$. The constant $C_\chi$ corresponds to the energy of the solution, via eq. \eqref{massHP}, with $C_\varphi$ replaced by $\frac{1}{\sqrt{2}}C_\chi$, due to \eqref{chieqphi}. The temperature is independent of the energy for these solutions, which means that they describe a system with Hagedorn thermodynamics.

The explicit solution for $\chi(r)$ allows us to evaluate its free energy. Using \eqref{chieqphi} and \eqref{chisol} in \eqref{Sphichi}, we find
\begin{equation}
\label{Fd6}
    \begin{split}
        F=\frac{I_d}{\beta}=\frac{1152}{5\kappa^2}\omega_5\frac{\alpha'^2}{16\pi G_N}~,
    \end{split}
\end{equation}
which is independent of the parameter $C_\chi$, and thus of the mass. This is a manifestation of the conformal symmetry of the classical action \eqref{Sphichi} discussed below \eqref{eom6}.

It is interesting to note that despite the thermodynamics being Hagedorn, the free energy is non-zero. This is necessary for the classical solution to be reliable, and quantum corrections to be suppressed. From \eqref{Fd6}, quantum corrections are suppressed provided that 
\begin{equation}
    \begin{split}
        g_s\ll 1~,
    \end{split}
\end{equation}
which is independent of $C_\chi$ and only requires the theory to be weakly coupled. The solution \eqref{chisol} can furthermore only be trusted in the regime where classical corrections to the action \eqref{Sphichi} can be neglected, i.e. for $\chi(r)\ll1$. From \eqref{chizz}, we find the constraint
\begin{equation}
\label{regval}
    \begin{split}
        \chi(0)\ll1, \;\;{\rm or\; equivalently}\;\;C_\chi\gg \alpha'^2~.
    \end{split}
\end{equation}
As discussed above, beyond this regime one has to add to the action \eqref{Sphichi} terms of higher order in $\chi$, $\varphi$ and derivatives. In terms of the mass $M$ of the solution, the condition \eqref{regval} is equivalent to $M\gg M_{\mathrm{corr}}$, where $M_{\mathrm{corr}}$ is given by \eqref{mcorr}.

\subsection{The limit $d\to6$ of HP}
\label{sec:d=6}

The solution of the effective Lagrangian \eqref{Sphichi} studied in the previous subsection seems a priori distinct from the Horowitz-Polchinski solution reviewed in section \ref{sec:hpsol}. Indeed, the former only exists for $d=6$, $m_\infty=0$, while the latter is non-trivial for $2<d<6$ and $m_\infty>0$. In this subsection we show that the solution of section \ref{sec:classical sols} can be obtained from that of section \ref{sec:hpsol} via a double scaling limit, $m^2_\infty,6-d\to0$ with the ratio $m^2_\infty/(6-d)$ held fixed.

In section \ref{sec:hpsol} we have seen that as the dimension $d$ increases, with $m_\infty$ held fixed, the HP solution becomes less controlled. For example, looking back at figure \ref{chiphi0figs}, we see that $\hat{\chi}(0),\hat{\varphi}(0)$ appear to grow without bound as $d$ approaches some critical value. We will next show that this critical value is $d=6$, and $\hat{\chi}(0),\hat{\varphi}(0)$ have a simple pole there. 

To do that, it is convenient to perform a different rescaling of the fields $\chi(x),\varphi(x)$ than that given in \eqref{rescaled}, 
	\begin{equation}
	\label{rescalzeta}
	\begin{split}
 	x&=\tilde{x}\frac{\sqrt{\zeta}}{m_{\infty}} \ ,\\
 	\chi (x)&=\frac{\alpha' \sqrt{2}}{\kappa}\frac{m^2_{\infty}}{\zeta}\tilde{\chi}(\tilde{x})\ ,\\
 	\varphi (x)&=\frac{m^2_{\infty}\alpha'}{\kappa \zeta}\tilde{\varphi}(\tilde{x})\ .
	\end{split}
	\end{equation}
 $\zeta$ is an arbitrary parameter, which depends on the choice of boundary conditions. Of course, since the solution of the original equations of motion \eqref{eom0} is unique, $\zeta$ is a redundant parameter, in the sense that quantities like $\tilde\varphi(0)/\zeta$, $\tilde\chi(0)/\zeta$ are independent of $\zeta$. In \eqref{rescaled} we took $\zeta=1$. Here, we will instead take $\tilde{\varphi}(0)=-1$, for any $d$, and compute $\zeta$ for this choice. 

The equations of motion \eqref{eom0}, written in terms of the rescaled fields \eqref{rescalzeta} are given by
	\begin{equation}
	\label{dezeta}
	\begin{split}
 	\tilde{\chi}''+\frac{d-1}{\tilde{r}}\tilde{\chi}' &=(\zeta+\tilde{\varphi})\tilde{\chi} \ ,\\
 	\tilde{\varphi}''+\frac{d-1}{\tilde{r}}\tilde{\varphi}' &=\tilde{\chi}^2\ .
	\end{split}
	\end{equation}
In Appendix \ref{zetad6}, we show that the HP solutions satisfy
	\begin{equation}
	\label{par}
	\begin{split}
 	\frac{1}{\zeta}\left(\frac{\int d\tilde{r}\ \tilde{r}^{d-1} \tilde{\varphi}(\tilde{r})(\tilde{\chi}(\tilde{r}))^2}{\int d\tilde{r}\ \tilde{r}^{d-1}(\tilde{\chi}(\tilde{r}))^2}\right)=\frac{4}{d-6}\ .
	\end{split}
	\end{equation}
	Since $\tilde{\varphi}(\tilde{r})$ is a negative, monotonically increasing function and $\tilde{\varphi}(0)=-1$, it follows that
	\begin{equation}
	    \begin{split}
	        0<\zeta<\frac{6-d}{4}~.
	    \end{split}
	\end{equation}
Hence, $\zeta\to0$ as $6-d\to0$. This establishes that $\varphi(0)$, which according to \eqref{rescalzeta} is proportional to $\tilde\varphi(0)/\zeta$, diverges as $d\to 6$, in agreement with what we found numerically in figure \ref{chiphi0figs}. A similar conclusion holds for $\chi(0)$.  

In the limit $\zeta\to 0$, the equations of motion \eqref{dezeta} for $\tilde{\chi},\tilde{\varphi}$ become the same as the equations of motion \eqref{eom6}, up to a finite rescaling of the fields. Therefore, if we are looking for normalizable solutions of \eqref{dezeta}, we can borrow the $d=6$ solution we found in the previous section, written in \eqref{chisol}, which after appropriately rescaling and imposing the boundary condition $\tilde\varphi(0)=-1$, is given by
	\begin{equation}
	\label{dschi}
	\begin{split}
 	\tilde\chi(\tilde{r})=-\tilde\varphi(\tilde{r})=\frac{576}{(24+\tilde{r}^2)^2}\ .
	\end{split}
	\end{equation}
	 Plugging this solution back into \eqref{par}, we conclude that $\zeta$ is given by
	\begin{equation}
\label{zeta6d}
    \begin{split}
        \zeta=\frac{6-d}{40}+{\mathcal{O}}\left((6-d)^2\right),
    \end{split}
\end{equation}	
and hence goes to zero linearly with $6-d$, as $6-d\to0$.

The above analysis explains the results of figure \ref{chiphi0figs}. Holding $m_\infty$ fixed and sending $d\to 6$, we see from equation \eqref{rescalzeta} that $\chi(0)$, $\varphi(0)$ diverge in the limit. We can keep them finite by considering the double-scaling limit  
\begin{equation}
\label{dslim}
    \begin{split}
    m_\infty\to0~,~~~d\to6~,~~~\frac{m^2_\infty}{6-d}=\Lambda^2\;\;\mathrm{fixed}~.
    \end{split}
\end{equation}
In this limit, $\chi(r)$, $\varphi(r)$ approach a solution of \eqref{eom6}, with $\varphi(0)=-\frac{40}{\kappa} \Lambda^2\alpha'$. Since this solution is normalizable, by construction, it is guaranteed to satisfy \eqref{chieqphi}. 
The double scaling parameter $\Lambda$ \eqref{dslim} can be written in terms of the parameters of section \ref{sec:classical sols} by noting that there one finds (from \eqref{chieqphi} and \eqref{chizz}) that $\varphi(0)=-\frac{576\sqrt{2}}{\kappa^2}\frac{\alpha'^2}{C_\chi}$. Comparing these two forms of $\varphi(0)$, we find 
\begin{equation}
\begin{split}
     \Lambda^2=\frac{72\sqrt{2}}{5\kappa}\frac{\alpha'}{C_\chi}~.
\end{split}
\end{equation}
One can interpret the double scaling parameter $\Lambda$ as giving the ``size'', or radial extent, of the solution. Indeed, \eqref{rescalzeta} suggests that the latter is given by
	\begin{equation}
	\label{lophi}
	\begin{split}
 	l&=\sqrt{\frac{\zeta}{m_{\infty}^2}}\sim\frac{1}{\Lambda}\;.
	\end{split}
	\end{equation}

The thermodynamic properties of the $d=6$ solution can be obtained by taking the double scaling limit of the corresponding properties of the Horowitz-Polchinski solution. Using \eqref{rescalzeta}, \eqref{dschi}, \eqref{zeta6d}, and \eqref{dslim} in \eqref{massHP} and \eqref{entropyHP}, we find
	\begin{equation}
	\label{SMds}
	\begin{split}
 	S=\beta_H M=\frac{576}{5\kappa}\omega_5\frac{\alpha'\beta_H}{16\pi G_N}\frac{1}{\Lambda^2}\;.
	\end{split}
	\end{equation}
Thus, $\Lambda$ can be thought of as parametrizing the mass of the solution. As mentioned above, $\chi(0), -\varphi(0)\propto\Lambda^2$. The temperature of the solution is the Hagedorn temperature, $T=T_H$, independently of the mass, as appropriate for Hagedorn thermodynamics \eqref{SMds}. 

For the free energy \eqref{FHP}, we find in the double scaling limit
	\begin{equation}
	    \begin{split}
	        F=\frac{1152}{5\kappa^2}\omega_5\frac{\alpha'^2}{16\pi G_N}\;.
	    \end{split}
	\end{equation}
It is constant, independent of the mass, and agrees with \eqref{Fd6}. Note that $-\beta F$ agrees with the constant $\xi_6$ in \eqref{SHPcorr} in the double-scaling limit, as expected from the thermodynamic relation $S(M)=\beta (M-F)$.
	
Combining \eqref{lophi} with \eqref{SMds}, we get the relation between the size and mass,
	\begin{equation}
	\begin{split}
 	l&=\sqrt{\frac{\kappa\pi G_N}{288\omega_5\alpha'}M}\ .
	\end{split}
	\end{equation}

\section{Worldsheet analysis and enhanced symmetry}
\label{sec:ws}

In section \ref{sec:classical sols} we found that the normalizable solution at $T=T_H$ in $d=6$ dimensions satisfies the relation \eqref{chieqphi}. From the point of view of the spacetime EFT \eqref{Sphichi}, this relation seems accidental. In this section we will argue that it is actually a consequence of an enhanced symmetry of the underlying worldsheet CFT.  

As we reviewed in section \ref{sec:hpsol}, the starting point of the HP analysis is string theory on $\mathbb{R}^d\times \mathbb{S}^1$ (times a compact space, that is a spectator in the discussion). At the Hagedorn temperature, the radius of the $\mathbb{S}^1$ is $R_H$, given in \eqref{tthh} for the bosonic and type II string. 

At this value of the radius, the $U(1)_L\times U(1)_R$ symmetry corresponding to momentum and winding on the Euclidean time circle is enhanced to $SU(2)_L\times SU(2)_R$, which turns out to play an important role in the analysis. To see this, we next discuss separately the two cases (bosonic and type II).

\subsection{Bosonic string}

Consider first the bosonic string. On a circle of radius $R$, the spectrum of left and right-moving momenta is given by the standard formula
\begin{equation}
	\label{plr}
	\begin{split}
 	p_L &=\frac{n}{R}+\frac{wR}{\alpha'} \ ,\\
 	p_R &=\frac{n}{R}-\frac{wR}{\alpha'} \ ,
	\end{split}
	\end{equation}
where $n$, $w$ are the integer momentum and winding, respectively. 

For $R=l_s$, the operators 
\begin{equation}
\label{eplpr}
	\begin{split}
 	e^{ip_LX_L+ip_RX_R} 
	\end{split}
	\end{equation}
with $n=w=\pm 1$ are holomorphic operators of dimension $h=1$,
\begin{equation}
\label{jpm}
	\begin{split}
 	J^\pm=e^{\pm2iX_L/l_s} .
	\end{split}
	\end{equation}
Together with the current $J^3=\frac{i}{l_s}\partial X_L$, these operators form a level one $SU(2)_L$ current algebra (see e.g. \cite{DiFrancesco:1997nk}). Similarly, the operators \eqref{eplpr} with $n=-w=\pm 1$ give rise to anti-holomorphic dimension one operators $\bar J^\pm$, which together with $\bar J^3=\frac{i}{l_s}\bar\partial X_R$ form a level one $SU(2)_R$ current algebra.

We are interested in the theory with $R=R_H=2l_s$, \eqref{tthh}, which can be thought of as a $\mathbb{Z}_2$ orbifold of the one with $R=l_s$. This orbifold can be described by adding to the theory with $R=l_s$ the operator \eqref{eplpr} with $n=\frac{1}{2}$, $w=0$. Mutual locality of the original operators \eqref{plr}, \eqref{eplpr} with this operator leads to the requirement $w\in 2\mathbb{Z}$. This is the untwisted sector of the orbifold. The twisted sector adds operators with $n\in \mathbb{Z}+\frac12$. Altogether, we have the spectrum of the theory with double the radius.   

The $SU(2)$ currents \eqref{jpm} have odd winding in the theory with $R=l_s$, and thus are not in the spectrum of the orbifolded theory. However, as the above presentation makes clear, they are mutually local w.r.t. the operators in the untwisted sector of the orbifold, and have a square root branch cut with those in the twisted sector. This is just another way of saying that they are odd under the $\mathbb{Z}_2$.

Their role in the theory is similar to that of supercurrents in a superconformal field theory, that are mutually local with NS sector operators, and have a square root branch cut with Ramond sector ones. In that case, the supercurrents themselves are not in the spectrum, but products of an even number of them are in the spectrum. Furthermore, if we deform the Lagrangian by NS sector operators only, the supercurrents help analyze the resulting deformation.

In our case, the operators \eqref{jpm} are similarly useful, since from the worldsheet point of view, turning on the spacetime fields $\varphi$, $\chi$ in $\eqref{Sphichi}$, corresponds to adding to the worldsheet Lagrangian the terms\footnote{We will discuss the relative normalization of the different terms below.}  
\begin{equation}
\label{delll}
	\begin{split}
 	\delta L=\frac{2}{\alpha'}\varphi(r)\partial X\bar\partial X+\frac{1}{\sqrt2}\chi(r)J^+\bar J^-+\frac{1}{\sqrt2}\chi^*(r)J^-\bar J^+\ .
	\end{split}
	\end{equation}
All the terms in \eqref{delll} belong to the untwisted sector of the above $\mathbb{Z}_2$ orbifold. Thus, they are mutually local w.r.t. the currents \eqref{jpm} and their right-moving analogs. Hence, we can use the $SU(2)_L\times SU(2)_R$ current algebra to study the perturbed theory. 

In terms of $SU(2)$ currents, \eqref{delll} can be written as 
\begin{equation}
\label{delcur}
	\begin{split}
 	\delta L=-2\varphi(r)J^3\bar J^3+\frac{1}{\sqrt2}\chi(r)J^+\bar J^-+\frac{1}{\sqrt2}\chi^*(r)J^-\bar J^+\ .
	\end{split}
	\end{equation}
and if \eqref{chieqphi} is satisfied, 
\begin{equation}
\label{thirr}
	\begin{split}
 	\delta L=-\varphi(r)k_{ab}J^a\bar J^b\ ,
	\end{split}
	\end{equation}
where $k_{33}=2$, $k_{+-}=k_{-+}=1$. 

The combination $k_{ab}J^a\bar J^b$ in \eqref{thirr} is nothing but the non-abelian Thirring perturbation for $SU(2)$ \cite{Dashen:1973nhu,Dashen:1974hp,Kutasov:1989dt}, with $k_{ab}$ the inverse of the matrix of two point functions $k^{ab}=\langle J^a J^b\rangle$. On its own, it breaks conformal invariance and generates an RG flow. It also breaks the $SU(2)_L\times SU(2)_R$ symmetry of the undeformed theory to the diagonal $SU(2)$. In our case, \eqref{thirr}, the coupling is itself a function of the radial coordinate on $\mathbb{R}^6$. From the worldsheet point of view, its $r$ dependence is fixed by the requirement that the full theory remains conformally invariant. As usual (see e.g. \cite{Polchinski:1998rq}), this is the same as the requirement that the spacetime equations of motion of the corresponding string theory are satisfied.

We now see that configurations which satisfy \eqref{chieqphi} are special from the worldsheet point of view, in that they break the $SU(2)_L\times SU(2)_R$ symmetry of $\mathbb{R}^6\times \mathbb{S}^1$ at the Hagedorn radius to the diagonal $SU(2)$. This is in contrast to what happens for general temperatures (e.g. for the HP solution, and for EBH's with $T<T_H$), where the symmetry breaking pattern is $U(1)_L\times U(1)_R\to U(1)_{\rm diagonal}$. 

We note in passing that while the deformed theory \eqref{thirr} is a CFT with a conserved $SU(2)$ current, 
\begin{equation}
    \begin{split}
        \bar\partial J^a+\partial\bar J^a=0~,
    \end{split}
\end{equation}
it does not have separately conserved holomorphic and anti-holomorphic currents. This is due to the non-compactness of the theory.

In our discussion above we assumed that the relative normalization of $\varphi$ and $\chi$ given in \eqref{delcur} is the same as that in the spacetime effective action \eqref{Sphichi}. In order to show that this is indeed correct, we note that the fact that in the spacetime analysis of section \ref{sec:hpsol} the kinetic terms of $\varphi$ and $\chi$ have the same normalization means that the worldsheet operators multiplying $\varphi$ and $\chi$ in \eqref{delcur} must have the same two point function. 

This is indeed the case: $\varphi(r)$ multiplies the operator $-2J^3\bar J^3$, whose two point function is one, and if we write $\chi=\chi_1+i\chi_2$, the field $\chi_1(r)$ multiplies the operator $(J^+\bar J^-+J^-\bar J^+)/\sqrt2$, whose two point function is also equal to one. Thus, the relative normalization of $\varphi$ and $\chi$ in \eqref{delll}, \eqref{delcur} is the same as that in the spacetime action \eqref{Sphichi}. 

\subsection{Type II superstring}

In this case, $R_H=\sqrt2l_s$ \eqref{tthh}. The worldsheet theory on the $\mathbb{S}^1$ is now an $N=1$ superconformal field theory of a scalar field $X$ living on a circle of radius $R_H$, a real left-moving fermion $\psi_X$, and a real right-moving fermion $\bar\psi_X$. These fields are related by worldsheet supersymmetry, under which they form a real superfield $X+\theta\psi_X+\bar\theta\bar\psi_X+\cdots$.

As is well known \cite{DiFrancesco:1997nk}, precisely when the radius of the $\mathbb{S}^1$ is equal to $\sqrt 2 l_s$ (or its T-dual, $l_s/\sqrt2$), the CFT of $X$ is equivalent to that of two real left-moving fermions, $(\psi_1,\psi_2)$, and two right-moving ones, $(\bar\psi_1,\bar\psi_2)$. Denoting the fermions $\psi_X$, $\bar\psi_X$  by $\psi_3$, $\bar\psi_3$, respectively, the SCFT on $\mathbb{S}^1$ becomes a theory of three (left and right-moving) real free fermions $\psi^a$, $\bar\psi^a$, $a=1,2,3$. 

The $N=1$ superconformal generators are given in terms of the fermions by 
\begin{equation}
	\label{Glr}
	\begin{split}
 	G &=\psi_1\psi_2\psi_3 \ ,\\
 	\bar G &=\bar\psi_1\bar\psi_2\bar\psi_3 \ .
	\end{split}
	\end{equation}
The worldsheet theory has in this case an $SU(2)_L\times SU(2)_R$ supersymmetric affine Lie algebra. The $SU(2)_L$ supercurrents are given by $\Psi^a=\psi^a+\theta J^a$, with $a=1,2,3$. Using \eqref{Glr} one finds that $J^a\sim \epsilon^{abc}\psi^b\psi^c$, which forms a level two $SU(2)$ affine Lie algebra. Similar equations with $L\to R$ describe the $SU(2)_R$. Like in the bosonic theory, the left and right-moving $SU(2)$ currents are not in the spectrum  of the GSO projected theory, but products of even numbers of them are.

One can generalize the worldsheet analysis of the bosonic case to the type II one. The analogs of equations \eqref{delll} -- \eqref{thirr} for that case must preserve worldsheet SUSY, and are thus naturally written in terms of superfields. Denoting by $X^i$, $i=1,2,\cdots, 6$ the superfields $X^i=x^i+\theta\psi^i+\bar\theta\bar\psi^i+\cdots$, the analog of \eqref{thirr} for this case is 
\begin{equation}
\label{susythirr}
	\begin{split}
 	\delta L=-\int d^2\theta\varphi(r)k_{ab}\Psi^a\bar \Psi^b\ ,
	\end{split}
	\end{equation}
where $r$ is the radial part of the superfields $X^i$, $r^2=X^iX^i$. Like in the bosonic case, this deformation (which describes the case where $\varphi$ and $\chi$ satisfy the relation \eqref{chieqphi}) breaks the $SU(2)_L\times SU(2)_R$ symmetry of the asymptotic background to the diagonal $SU(2)$.

\subsection{Worldsheet dynamics}

To recapitulate, we showed that the $d=6$ solution of the EFT which we found in section \ref{sec:classical sols} corresponds from the worldsheet perspective to a CFT on $\mathbb{R}^6\times \mathbb{S}^1$ in the presence of a non-abelian Thirring perturbation with a coupling that depends on the radial coordinate on $\mathbb{R}^6$, $\varphi(r)$. 

From this perspective, the form of $\varphi(r)$ is obtained by setting to zero the $\beta$-function of the perturbed model \eqref{thirr}, \eqref{susythirr}, for the bosonic and type II string, respectively. The exact form of this $\beta$-function is not known, however we can study it perturbatively in $\varphi$. To do that, recall that if we perturb a CFT by adding to the Lagrangian 
\begin{equation}
\label{pertbeta}
	\begin{split}
 	\delta L=\sum_i\lambda_i\Phi^i\ ,
	\end{split}
	\end{equation}
where $\Phi_i$ are conformal primaries of scaling dimension $h=\bar h=\Delta_i$ and $\lambda_i$ are the corresponding couplings, the $\beta$ functions for $\lambda_i$, $\beta_i$, are given by \cite{Zamolodchikov:1987ti}
\begin{equation}
\label{formbeta}
	\begin{split}
 	\beta_i=-(1-\Delta_i)\lambda_i+C_{ijk}\lambda_j\lambda_k+O(\lambda^3)\ .
	\end{split}
	\end{equation}
Here, $C_{ijk}$ is proportional to the three point function 
\begin{equation}
\label{threept}
	\begin{split}
 	C^{ijk}\sim\langle\Phi^i\Phi^j\Phi^k\rangle \ .
	\end{split}
	\end{equation}
To apply \eqref{formbeta} to our case, we can Fourier expand the field $\varphi(x)$, 
\begin{equation}
\label{fourr}
	\begin{split}
 	\varphi(x)=\int d^dk\phi(k) e^{ik\cdot x}\ ,
	\end{split}
	\end{equation}
and view $\phi(k)$ as the couplings $\lambda_i$ in \eqref{pertbeta}, \eqref{formbeta}, with the index $i$ running over all momenta $k$. Plugging \eqref{fourr} into \eqref{thirr}
we see that the operators $\Phi^i$ are given in this case by 
$k_{ab}J^a\bar J^be^{ik\cdot x}$, and the dimensions $\Delta_i$ 
are given by $1+\frac{\alpha'}{4}k^2$. Thus, the linear term in 
the $\beta$-function for $\varphi(x)$ is $\beta(x)\sim-\nabla^2\varphi+ O(\varphi^2)$.

To compute the quadratic term in the $\beta$-function, we need to evaluate the three point function of the perturbing operator. 
In momentum space this three point function is given by $C\delta^6(k_1+k_2+k_3)$, with $C$ a calculable constant. In position space this corresponds to $\varphi^2(x)$. Putting the two contributions together, we find 
\begin{equation}
\label{fullbeta}
	\begin{split}
 	\beta(x)\sim-\nabla^2\varphi+ C\varphi^2(x)+O(\varphi^3)\ .
	\end{split}
	\end{equation}
The condition $\beta(x)=0$ is precisely the equation that we solved in section \ref{sec:classical sols}, \eqref{neweom}. Thus, when $\chi(0)$ is small, we can use the worldsheet Lagrangian, \eqref{thirr} (or \eqref{susythirr}), to compute observables in the CFT corresponding to the solution constructed in section \ref{sec:classical sols}.  

When $\chi(0)$ increases, we need to include higher order corrections to the $\beta$-function \eqref{fullbeta}. One issue that arises is whether when we go outside the regime of validity of the effective Lagrangian \eqref{Sphichi}, the worldsheet CFT still preserves the diagonal $SU(2)$ subgroup of $SU(2)_L\times SU(2)_R$ that is present at $r\to\infty$. It is in principle possible that higher order corrections break this symmetry to the $U(1)$ corresponding to translations around the circle, but we believe that this $SU(2)$ is an exact symmetry of the worldsheet CFT for all values of $\chi(0)$. This symmetry breaking pattern seems to be a property of the theory at the Hagedorn temperature, and not of a particular approximation to it. We will present later some further evidence for this claim.

From the worldsheet perspective, the question whether the $SU(2)$ is a symmetry of the full theory is essentially the question whether perturbing the CFT on $\mathbb{R}^6\times \mathbb{S}^1$ by the non-abelian Thirring interaction \eqref{thirr} (or \eqref{susythirr}) leads to a theory that preserves $SU(2)$. Since this $SU(2)$ is non-chiral, we do not expect any anomalies that break it in the quantum theory.

Another issue that arises is the following. The non-abelian Thirring interaction \eqref{thirr} couples the $\mathbb{S}^1$ and the $\mathbb{R}^6$. Thus, it is natural to expect that the metric on $\mathbb{R}^6$ backreacts to the non-zero $\varphi$. The leading backreaction can be calculated either from the spacetime action \eqref{fac}, or by following the logic of this subsection. 
We can add to the worldsheet Lagrangian\footnote{In the bosonic case. The situation in the superstring is similar.} the term 
\begin{equation}
\label{metr}
	\begin{split}
 	\delta L=h_{ij}(x)\partial x^i\bar\partial x^j~.
	\end{split}
	\end{equation}
The leading term in the $\beta$-function for $h_{ij}$ is $\beta_{ij}\sim-\nabla^2 h_{ij}$. The next term comes from \eqref{formbeta}, where we take the operator $\Phi^i$ to be the one in \eqref{metr}, and the other two, $\Phi^j$ and $\Phi^k$, to be those from \eqref{thirr}. 

To evaluate this term, we need to compute the three point function 
\begin{equation}
\label{threept}
	\begin{split}
 	\langle h_{ij}(x)\partial x^i\bar\partial x^j \varphi(x)k_{ab}J^a\bar J^b \varphi(x)k_{cd}J^c\bar J^d\rangle~.
	\end{split}
	\end{equation}
This gives a contribution to the $\beta$-function $\beta_{ij}$ that is quadratic in $\varphi$, 
\begin{equation}
\label{betaij}
	\begin{split}
 	\beta_{ij}\sim-\nabla^2 h_{ij}+C\partial_i\varphi\partial_j\varphi
	\end{split}
	\end{equation}
for some constant $C$. 

Since the perturbation \eqref{thirr} preserves $SO(6)$, the backreaction should too. Therefore, we can write $h_{ij}=\omega\delta_{ij}$. Plugging this into \eqref{betaij}, we find 
\begin{equation}
\label{betaomega}
	\begin{split}
 	\beta_{\omega}\sim-\nabla^2 \omega+C\left(\nabla\varphi\right)^2~.
	\end{split}
	\end{equation}
The backreacted metric\footnote{More precisely, the back-reaction involves a combination of $\omega$ and the dilaton $\phi_d$ \eqref{fac}.} is obtained by setting \eqref{betaomega} to zero, i.e. by solving $\nabla^2 \omega=C\left(\nabla\varphi\right)^2$.
We see that $\omega$ has scaling dimension four, and gives a subleading effect in the small $\varphi$ expansion. As mentioned above, there are other subleading effects, such as those that come from higher order contributions to the Lagrangian for $\chi$, $\varphi$, \eqref{Sphichi}.

It would be interesting to solve the localized non-abelian Thirring CFT described in this section exactly, i.e. beyond the small $\varphi$ approximation.

\section{More on $d=6$}
\label{sec:md6}

In the previous sections we presented evidence for the claim that conformal field theory on $\mathbb{R}^6\times \mathbb{S}^1$ has an exactly marginal deformation of the form \eqref{thirr} in the bosonic case, and \eqref{susythirr} in the $N=1$ superconformal one. This deformation is labeled by a single parameter, which can be taken to be $\varphi(0)$. Given this parameter, the function $\varphi(r)$ is determined by the requirement of (super) conformal invariance. 

When $\varphi(0)\ll 1$, one can solve for $\varphi(r)$ by approximating the (S)CFT by an effective field theory, \eqref{Sphichi} with $m_\infty=0$. The equations of motion of this EFT \eqref{eom6} reduce in this case to \eqref{neweom}, whose solution is \eqref{chisol}. For general $\varphi(0)$, the EFT fails at small $r$ and one needs to go back to the full problem \eqref{thirr}, \eqref{susythirr}. 

Of course, without solving the CFT of section \ref{sec:ws}, one can not be completely sure that the conformal manifold labeled by $\varphi(0)$ extends to finite values of this parameter, and if so, that it preserves the $SU(2)$ symmetry discussed in that section. As we explained above, in this paper we take the attitude that both of those features are true. The main reason for that is that the construction is in the spirit of the $\epsilon$ expansion, with $\varphi(0)$ playing the role of $\epsilon$. Thus, it is natural to expect that a CFT exists for finite $\varphi(0)$, and that it preserves $SU(2)$. Note that the issue of the existence of the background for finite $\varphi(0)$ in principle also exists for the HP solutions reviewed in section \ref{sec:hpsol}, and we are making the same assumption there.

In this section, we will assume that the background parametrized by $\varphi(0)$ exists, and explore some of its properties. In particular, we will propose that it describes, in a certain limit, highly excited fundamental strings, and admits an effective worldsheet decription in terms of an $SL(2,\mathbb{R})/U(1)$ cigar geometry. We will then comment on the relation between the $d=6$ solution and the EBH and free string solutions.

\subsection{The scaling limit $g_s\to 0$, $M/m_s$ fixed}

When one embeds the above (S)CFT in string theory, and views the $\mathbb{S}^1$ as Euclidean time, one can think of it as a Euclidean configuration, with mass 
\begin{equation}
\label{msa}
	\begin{split}
\frac{M}{m_s}\sim\frac{a}{g_s^2}~.	\end{split}
	\end{equation}
The constant $a$ is proportional to $1/\varphi(0)$. The CFT of section \ref{sec:ws} describes the region in parameter space where $a$ remains finite in the limit $g_s\to 0$. Within this region, the EFT corresponds to $a\gg 1$. 

Another regime of interest is the {\it double scaling limit} $a\to0$, $g_s\to 0$, with the ratio $M/m_s$ \eqref{msa} held fixed and large. This regime of energies corresponds to that of highly excited fundamental strings, and it is natural to ask whether there is an analog of the above localized non-abelian Thirring description of this region. In this section we would like to propose such a description.

The region in question corresponds to large $\varphi(0)$, $1/\varphi(0)\sim a\sim g_s^2$. It is important to emphasize that it is not described by the CFT of section \ref{sec:ws}, since $\varphi(0)$ depends on $g_s$ in this case.

Looking back at the solution \eqref{chisol}, we see that in this case $C_\chi\sim g_s^2$ (due to \eqref{chizz}). Thus, the asymptotic form of $\chi(r)$ (and $\varphi(r)$, \eqref{chieqphi}), given by \eqref{dsixas} gives a profile that vanishes, as $g_s\to 0$, for all finite $r$. $\chi(r)$ reaches a value of order one at $r\sim\sqrt g_s$. The form \eqref{dsixas}, \eqref{chisol} cannot be trusted beyond that point, but if we were to continue to this regime, significant deviations from the asymptotic form would only appear at $r\sim g_s$. In other words, the whole non-trivial structure of the solution is packed in this case into a small region in $r$, the near-horizon region. 

A useful analogy is to Euclidean Schwarzschild black holes at large $d$. As $d\to\infty$, the Schwarzschild solution becomes trivial for all $r$ except for a small region around $r=R_{sch}$. In that case, it was found in \cite{Soda:1993xc,Emparan:2013xia,Chen:2021emg} that the near-horizon behavior is described by the two dimensional $SL(2,\mathbb{R})/U(1)$ black hole. 

We would like to do the same thing here. In our case, the problem cannot be addressed purely in the EFT, so we will make a proposal and present some evidence for its validity, leaving a more detailed analysis to future work.   

In order to study the near-horizon region for large $\chi(0)$, it is useful to integrate out the angular degrees of freedom on $\mathbb{R}^6$, and study the dynamics of the s-waves. This is analogous to what one does in the large $d$ analysis of \cite{Soda:1993xc,Emparan:2013xia,Chen:2021emg}. In that case, the $SL(2,\mathbb{R})/U(1)$ black hole is obtained after reducing the spacetime fields on the sphere. The reduction gives rise to the dilaton in the resulting two dimensional background. In our case, it is reasonable to integrate out the angular degrees of freedom, since in the near-horizon region the radius of the sphere is very small.

The resulting near-horizon background is a two dimensional sigma model for the radial coordinate and Euclidean time, that should look like a semi-infinite cigar (in the scaling limit). A natural candidate for that CFT is the $SL(2,\mathbb{R})/U(1)$ black hole at level $k=4$ in the bosonic string, and $k=2$ in the superstring. 

Both backgrounds are obtained by starting with a level four bosonic $SL(2,\mathbb{R})$ CFT. In the bosonic string we mod out by the timelike $U(1)$ subgroup, while in the superstring we add three free fermions that give a level $-2$ $SL(2,\mathbb{R})$ affine Lie algebra, for a total level $4-2=2$, and mod out by the corresponding timelike $U(1)$.

The level of $SL(2,\mathbb{R})$ is obtained in both cases by matching the asymptotic radius of the $\mathbb{S}^1$ as $r\to\infty$, with the asymptotic radius of the $SL(2,\mathbb{R})/U(1)$ cigar. This matching relies on the fact that in the scaling limit described above, the radius of the $\mathbb{S}^1$, which is controlled by $\varphi$, essentially does not change as $r$ varies from infinity to the near-horizon region. 

The scaling limit above contains a parameter, $m_s/M$, the mass of the solution in string units. We expect this parameter to correspond in the $SL(2,\mathbb{R})/U(1)$ cigar to the string coupling squared at the tip, as is familiar from studies of this background in the context of black hole physics (see e.g. \cite{Maldacena:1997cg}). Note that the asymptotic radius of the cigar is fixed, but the mass $M$ can vary, as expected for a system with Hagedorn thermodynamics. The condition $M\gg m_s$ is then the requirement that string loop corrections in this background are small. 

A test of this proposal is the following. We argued above that the $T=T_H$ solutions exhibit the symmetry breaking pattern 
\begin{equation}
\label{hagsym}
	\begin{split}
 	SU(2)_L\times SU(2)_R\to SU(2)_{\rm diagonal}\;.
	\end{split}
	\end{equation}
If the near-horizon geometry  in the scaling limit really is $SL(2,\mathbb{R})/U(1)$, it should also exhibit this symmetry breaking pattern. Interestingly, precisely for the values of $k$ found above, it is known that this is indeed the case. 

In particular, in the superstring, the $SL(2,\mathbb{R})/U(1)$ CFT with $k=2$ exhibits an enhanced symmetry, with the symmetry breaking pattern \eqref{hagsym}. In the context of Double Scaled Little String Theory \cite{Giveon:1999px, Giveon:1999tq}, this fact has a geometric interpretation in terms of fivebrane physics. 

The $SU(2)_L\times SU(2)_R$ is in that case interpreted as the $SO(4)$ symmetry of rotation about $k=2$ coincident fivebranes \cite{Callan:1991dj,Aharony:1998ub}, while the unbroken symmetry $SU(2)_{\rm diagonal}$
in \eqref{hagsym} is the residual $SO(3)$ symmetry of a configuration where the two fivebranes have been separated by a small amount \cite{Murthy:2003es}. The analog of the scaling limit discussed above is sending the distance between the fivebranes and $g_s$ to zero, while holding the mass of a D-string stretched between the fivebranes fixed and large in string units. 

The fact that for $k=4\; (2)$ in the bosonic string (superstring) the $SL(2,\mathbb{R})/U(1)$  CFT has the same symmetry enhancement that we found in sections \ref{sec:d6}, \ref{sec:ws}, is a non-trivial test of the proposal that it is the near-horizon geometry of these solutions in the scaling limit. 

As another check of this picture, we can compare the central charges of the worldsheet theories at large $r$ and in the near-horizon region. The large $r$ theory is conformal field theory on $\mathbb{R}^6\times \mathbb{S}^1$. Its central charge is 
\begin{equation}
\label{centch}
	\begin{split}
 	c^{\rm bosonic}=6+1=7,\;\; c^{\rm type\; II}=(6+1)\times\frac32=\frac{21}{2}~.
	\end{split}
	\end{equation}
The near-horizon $SL(2,\mathbb{R})/U(1)$ (S)CFT proposed above has central charge 
\begin{equation}
\label{centsltwo}
	\begin{split}
 	c^{\rm bosonic}=5,\;\; c^{\rm type\; II}=6~.
	\end{split}
	\end{equation}
The central charges \eqref{centsltwo} are smaller than the corresponding ones in \eqref{centch}. This agrees with the fact that the former are obtained from the latter by integrating out some fields (the angular degrees of freedom). 

\begin{figure}[t]
\centering
 \includegraphics[width=0.6\textwidth]{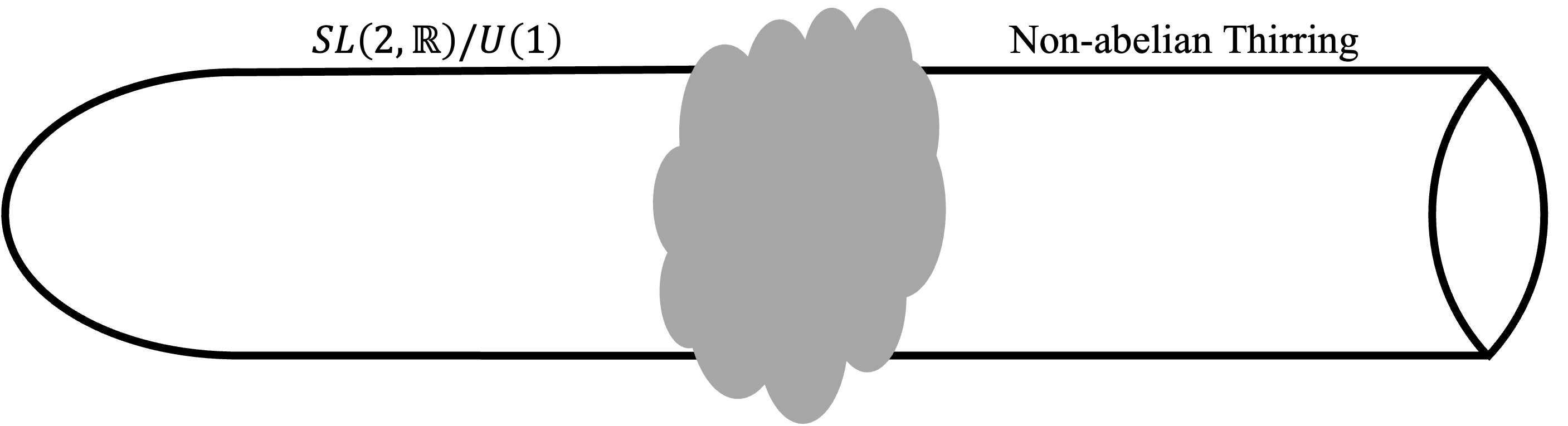}
 \caption{A background that interpolates between the two dimensional $SL(2,\mathbb{R})/U(1)$ EBH and $\mathbb{R}^6\times \mathbb{S}^1$ perturbed by the localized non-abelian Thirring interaction.}
\label{cigarR6S1}
\end{figure}

A natural question is what happens when we go beyond the double scaling limit, and consider solutions in which the asymptotic $g_s$ is non-zero. The rough picture we envision in this case is depicted in figure \ref{cigarR6S1}. The $SL(2,\mathbb{R})/U(1)$ EBH is now cut off at some finite value of the radial coordinate. At large $r$, the background approaches $\mathbb{R}^6\times \mathbb{S}^1$, perturbed by the non-abelian Thirring term \eqref{thirr}. 

The length of the two dimensional cigar throat depends on the mass of the solution $M$. Indeed, the string coupling at the tip of the $SL(2,\mathbb{R})/U(1)$ cigar is given by $g_{\rm tip}^2\sim m_s/M$ \cite{Maldacena:1997cg}. As one moves away from the tip, the string coupling decreases, and eventually it must approach the asymptotic value, $g_s$. The requirement that $g_{\rm tip}>g_s$ implies that $M<M_{\rm corr}$, where $M_{\rm corr}$ is the Horowitz-Polchinski correspondence mass \eqref{mcorr}. For the throat to be long requires $M\ll M_{\rm corr}$.

\subsection{Relation to Euclidean black hole and free string solutions}

\begin{figure}[t]
\centering
 \includegraphics[width=0.5\textwidth]{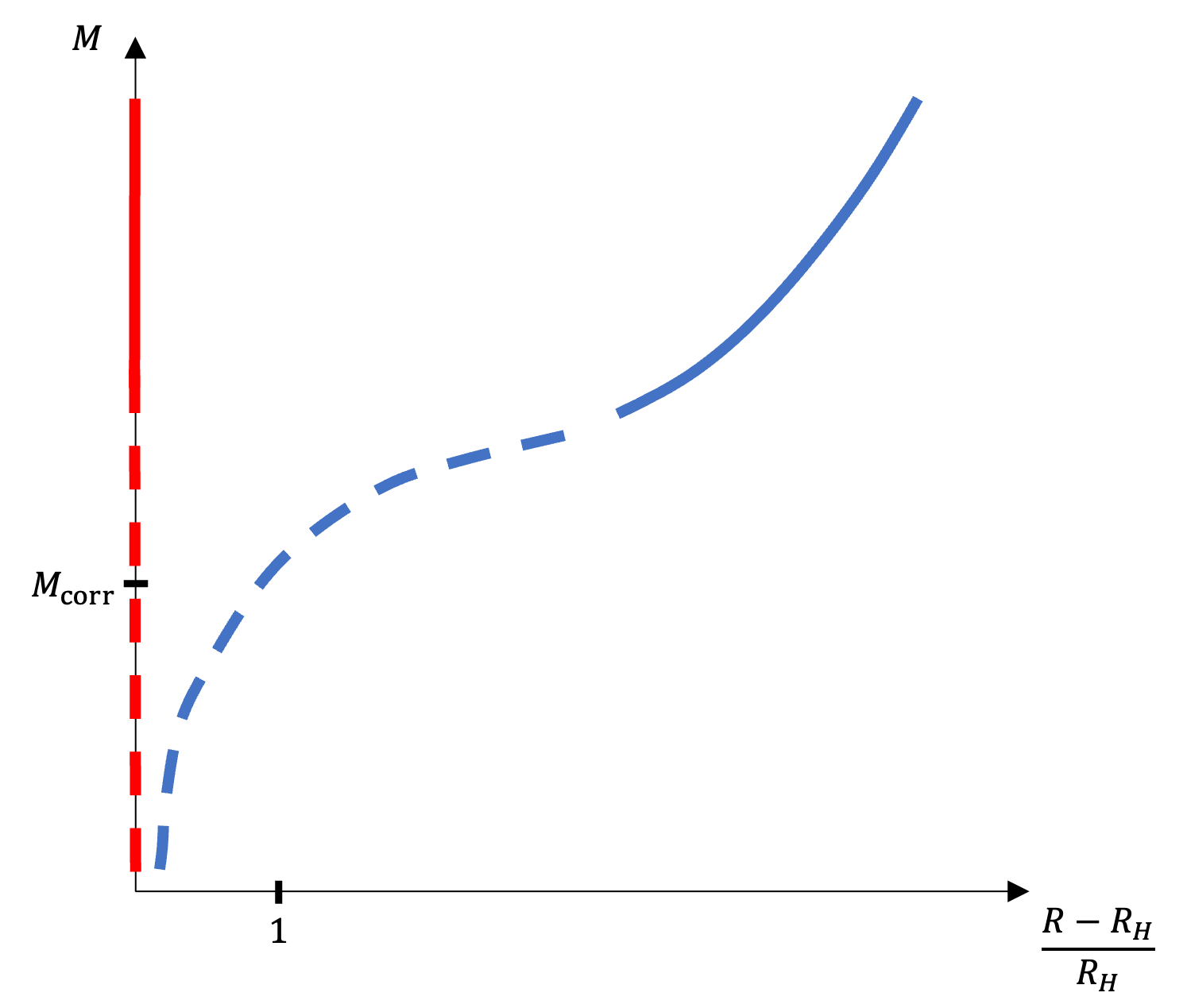}
 \caption{Mass vs temperature for the $d=6$ solutions. In red is the solution of sections \ref{sec:d6}, \ref{sec:ws}. In blue, the Euclidean Schwarzschild BH. Solid lines corresponds to regimes where the relevant EFT is valid, dashed lines are possible extrapolations.} 
\label{d6thermo}
\end{figure}

Another interesting question is the relation between the solution we constructed in this and the previous sections at the Hagedorn temperature, and the usual Euclidean Schwarzschild black hole. In figure \ref{d6thermo} we exhibit these solutions on a mass vs temperature plot.  

The red line represents the background we studied in this paper. The solid line corresponds to the region $\chi(0)\ll 1$, or $M\gg M_{\rm corr}$, which is well described by the EFT \eqref{Sphichi}. The dashed line is a continuation of this solution to all values of $M$. As mentioned above, it follows from the assumption that the solution can be continued to any value of $\chi(0)$, i.e. any value of $M$. If our conjecture from this section is correct, we have a partial understanding of the regime of $M$ corresponding to weakly coupled perturbative string states as well. In the figure, this is the region of small $M$. 

The blue line corresponds to Euclidean Schwarzschild black holes. For large $M$ they are well described by GR; hence the blue line is solid for $M\gg M_{\rm corr}$. As the mass decreases towards the correspondence mass, the Hawking temperatures is expected to approach $T_H$ \cite{Horowitz:1996nw}. A natural extrapolation to that regime, as well as to $M<M_{\rm corr}$ is given in the figure.  In the next section we will see that this extrapolation gives a natural continuation to dimensions $d<6$.

If this extrapolation is correct, the small Schwarzschild black hole approaches the background constructed in sections \ref{sec:d6}, \ref{sec:ws} as $T\to T_H$. In particular, if we consider the scaling limit $g_s\to 0$, $M/m_s$ fixed and large, figure \ref{d6thermo} suggests that the two solutions coincide. Thus, the Schwarzschild EBH approaches at small $M/M_{\rm corr}$ the $SL(2,\mathbb{R})/U(1)$ EBH of the same mass. 

For finite $g_s$ and $M\ll M_{\rm corr}$, we expect a picture similar to that of figure \ref{cigarR6S1}. The EBH develops (after reducing on the sphere) a long throat in which it looks like the two-dimensional $SL(2,\mathbb{R})/U(1)$ EBH. The difference between the red and blue curves in figure \ref{d6thermo} corresponds in this regime to the region beyond the cutoff in figure \ref{cigarR6S1}. For example, this is the region responsible for the small difference in asymptotic radii of Euclidean time in the two backgrounds. As $g_s\to 0$ (with fixed $M/m_s$) the cutoff goes to infinity, and the two backgrounds coincide. The resulting picture is reminiscent of \cite{Giveon:2006pr}, though some of the details are different.

So far, we discussed solutions that break the winding symmetry around Euclidean time. One can ask what is the relation of these solutions to the background $\mathbb{R}^6\times \mathbb{S}^1$, which does not break this symmetry. This background gives a contribution to the canonical partition sum that is of order $g_s^0$, and can be thought of as ${\rm Tr} e^{-\beta H}$ over the free string spectrum. Performing the trace explicitly, gives a mass that behaves as $R\to R_H$ like $M\sim (R-R_H)^{-1}$ (see e.g. \cite{Mertens:2015ola}). 

\begin{figure}[t]
\centering
 \includegraphics[width=0.5\textwidth]{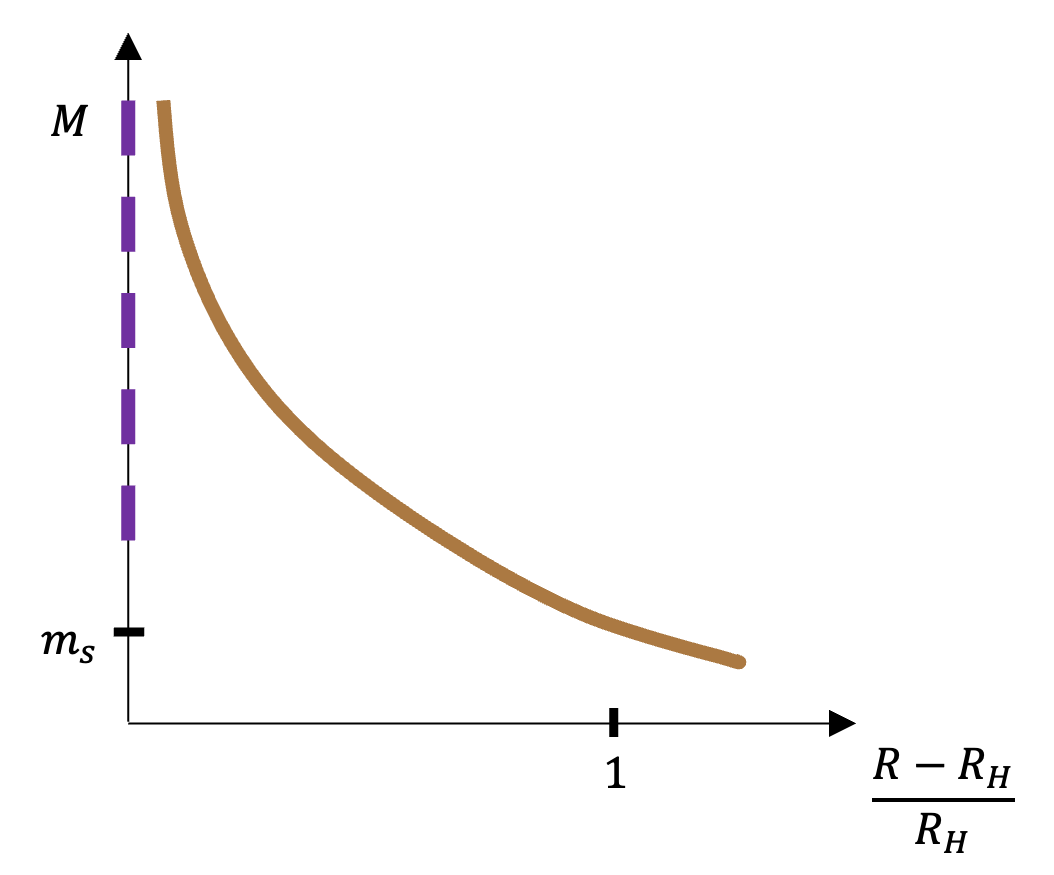}
 \caption{In the limit $g_s\to 0$, the winding violating solutions of figure \ref{d6thermo} approach the purple line. The brown curve describes the free string contribution $\mathbb{R}^6\times \mathbb{S}^1$. } 
\label{FSd6}
\end{figure}

Zooming in on masses of order $(g_s)^0$ in figure \ref{d6thermo}, and adding to it the free string curve gives rise to figure \ref{FSd6}. As discussed above, the red and blue curves of figure \ref{d6thermo} coincide in this region, and are thus jointly denoted by the purple line. The free string curve denoted in brown approaches it as $R\to R_H$ (i.e. for highly excited free strings). In this sense, the small black hole denoted by the purple line provides a good approximation to the free string, at least for $M\gg m_s$.   

It seems strange at first sight to compare the winding violating solutions described by the purple line and the winding preserving brown one. However, we note that the difference between the two is small in this case, in the sense that they coincide for finite $r$, and the winding violating is restricted to a region of size $g_s$. Also, describing fundamental strings as small black holes has a precedent in the study of BPS states (see e.g. \cite{Dabholkar:2004yr,Sen:1994eb,Sen:1995in}). Here we are discussing general states that contribute to the density of states \eqref{hagent}.

\subsection{A fivebrane analogue}

Finally, note that a similar picture to the one we propose for the EBH solution appears in a different context, related to fivebrane physics \cite{Maldacena:1997cg}. The near-extremal string-frame Euclidean solution for $k$ NS5-branes is given by \cite{Horowitz:1991cd}
\begin{equation}
    \begin{split}
        &ds^2=\left(1-\frac{r_0^2}{r^2}\right)d\tau^2+\left(1+\frac{k\alpha'}{r^2}\right)\left(\frac{dr^2}{1-\frac{r_0^2}{r^2}}+r^2 d\Omega_3^2\right)+dy_5^2~,
        \\
        & e^{2\phi_{10}}=g_s^2\left(1+\frac{k\alpha'}{r^2}\right)~,
    \end{split}
    \label{kns5metric}
\end{equation}
where $\phi_{10}$ is the ten-dimensional dilaton, and we omitted the background NS flux. The temperature of the solution is given by
\begin{equation}
    \begin{split}
        \beta=2\pi\sqrt{r_0^2+k\alpha'}~,
    \end{split}
    \label{kns5beta}
\end{equation}
and the energy density above the extremal value is given by
\begin{equation}
\begin{split}
    M&=\frac{m_s^8}{\left(2\pi\right)^5}\frac{r_0^2}{g_s^2}
    \\
    &=\frac{m_s^8}{\left(2\pi\right)^5g^2_s}\left(\frac{\beta^2}{(2\pi)^2}-k\alpha'\right).
    \end{split}
\end{equation}
Hence, the mass-temperature relation of this system is qualitatively similar to that of the EBH solution extrapolated to masses $M\lessapprox M_{\mathrm{corr}}$ as we proposed above (the blue line in figure \ref{d6thermo}), where the analogue of the Hagedorn temperature here is $\beta_H=2\pi\sqrt{k\alpha'}$.

As discussed in \cite{Maldacena:1997cg}, the limit $g_s\to0$ with $r_0/g_s$ held fixed decouples the asymptotically flat-space region, and zooms in the near-horizon region of \eqref{kns5metric}, where the theory admits a worldsheet description in terms of a $SL(2,\mathbb{R})_k/U(1)$ cigar CFT. The mass $M$ is held fixed and of order $(g_s)^0$ in this limit, and is related to the string coupling at the tip of the cigar. Furthermore, in this limit the temperature \eqref{kns5beta} is fixed at the ``Hagedorn temperature" $\beta_H=2\pi\sqrt{k\alpha'}$. 

Thus, the same features we expect for the ``small" EBH solution, and the emergence of the $SL(2,\mathbb{R})/U(1)$ cigar CFT, appear already in the context of near-extremal fivebrane solutions. The main difference is that in the context of fivebrane physics, the limit $k\to\infty$ allows the solutions \eqref{kns5metric} to be studied in supergravity, but such a limit does not exist for small EBHs.

\section{Discussion}
\label{sec:discu}

One of the problems left open by the analysis of the previous sections is what happens for $d\not=6$. In this section we will comment on this issue, leaving a more detailed discussion for future work.  

\begin{figure}[t!]
\centering
 \subfloat[\label{dl4}]{\includegraphics[width=0.3\textwidth]{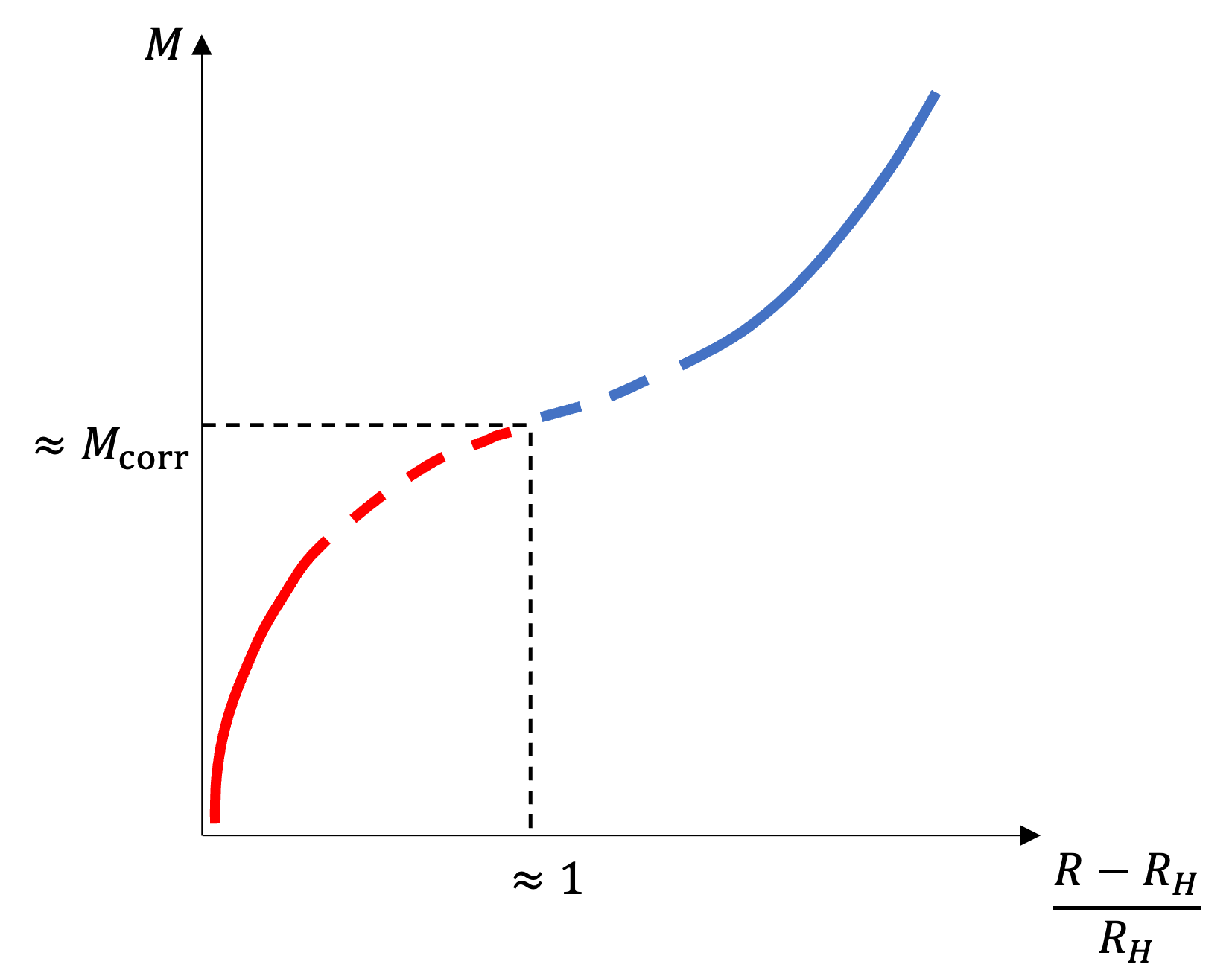}} 
 \subfloat[\label{d4}]{\includegraphics[width=0.3\textwidth]{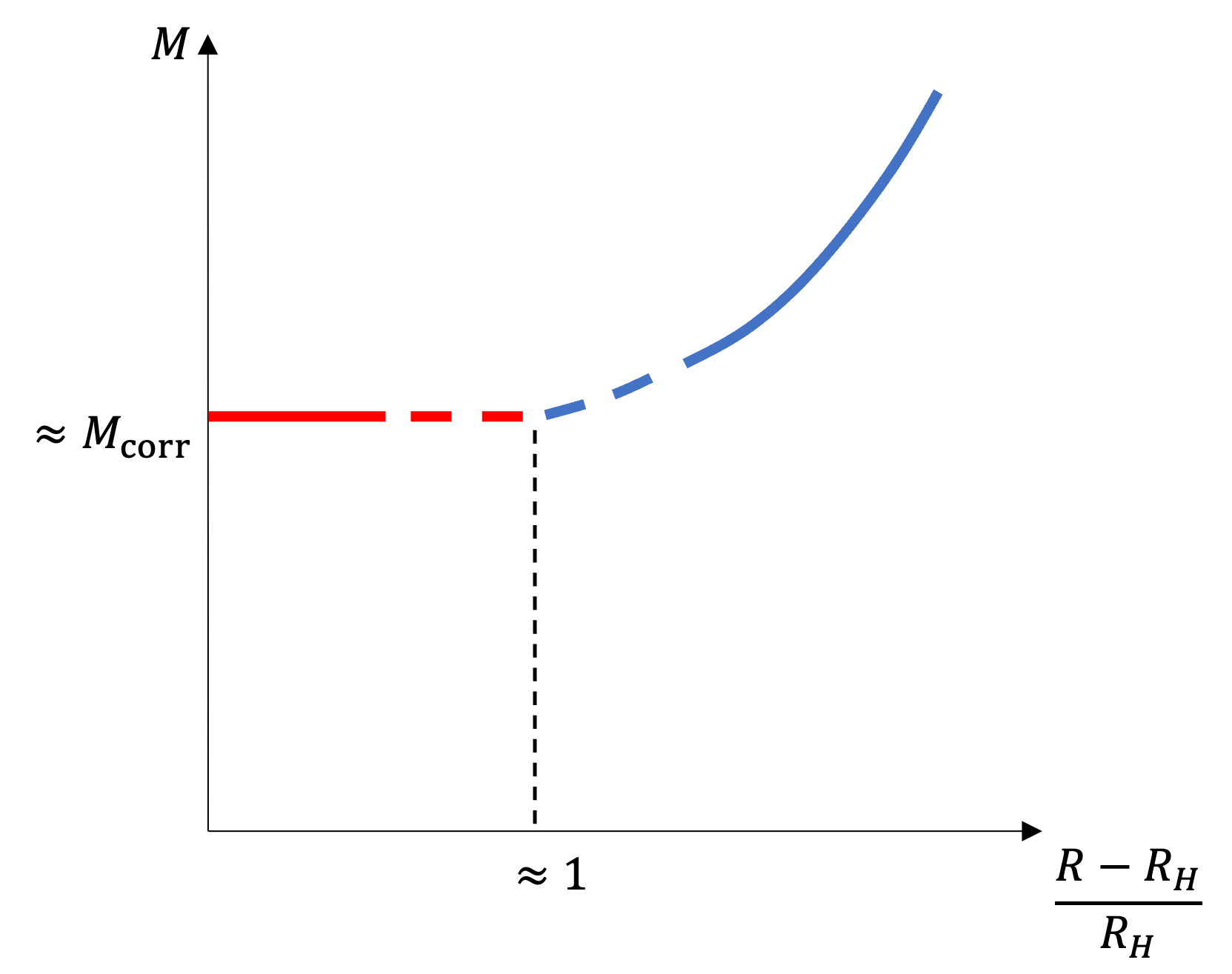}} 
 \subfloat[\label{4ldl6}]{\includegraphics[width=0.3\textwidth]{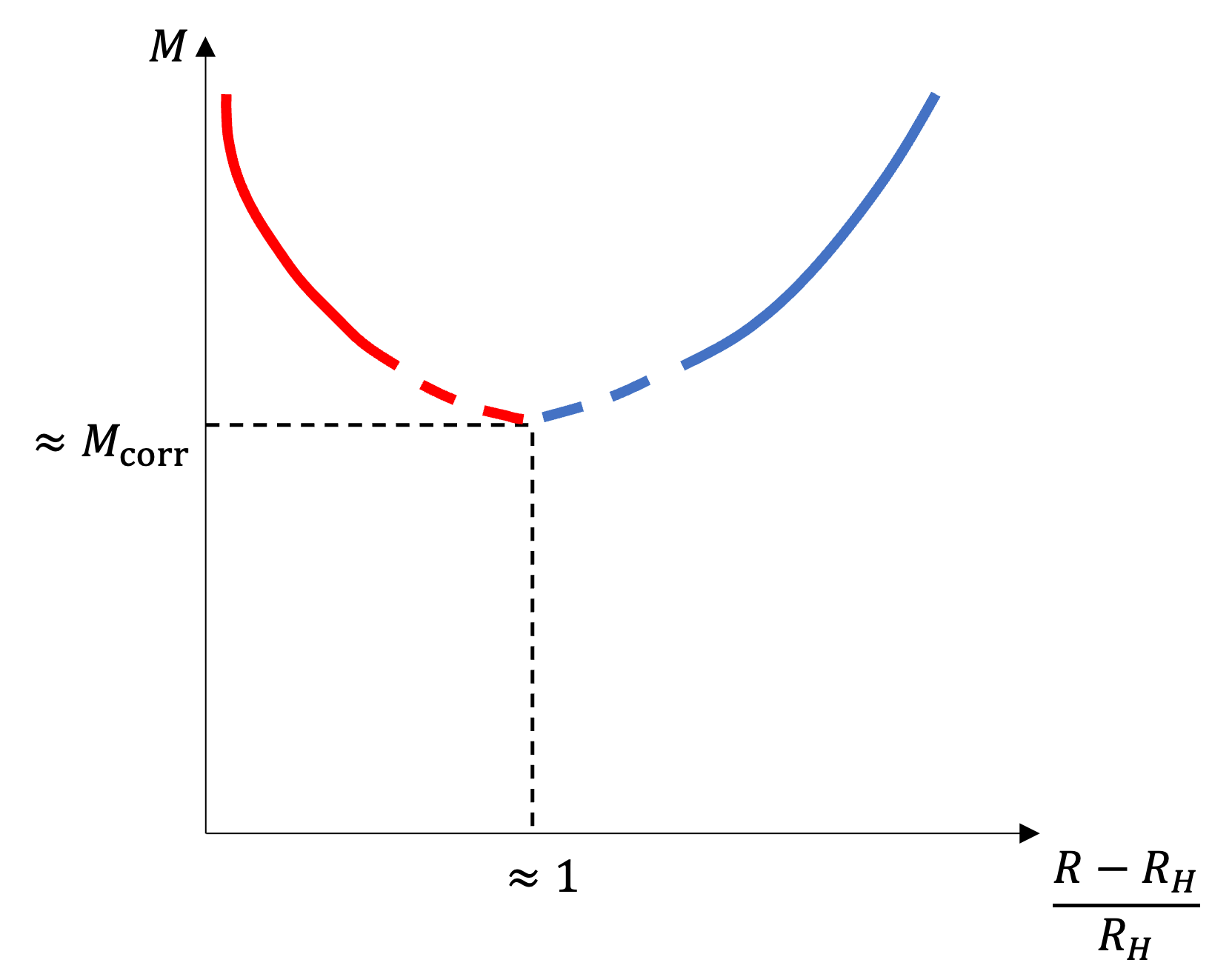}}
 \caption{Mass vs temperature of HP solutions (red) and EBH solutions (blue) for (a) $d<4$, (b) $d=4$ and (c) $4<d<6$.}
 \label{dneq6thermo}
\end{figure}

Consider first the case $d<6$. In this case, there are two solutions that are known to exist: the HP solution, that is reliably described by the EFT \eqref{Sphichi} for $m_\infty\ll m_s$ \eqref{gscor}, i.e. temperature close to the Hagedorn temperature, and the large EBH, which is reliably described by GR for temperature well below $T_H$. These solutions are described by the solid red and blue lines in figure \ref{dneq6thermo}.

It is natural to ask what happens to these solutions outside of the regime of validity of their respective EFT's. For example, the EBH corresponds to a line of classical solutions labeled by the Hawking temperature, or alternatively by the mass of the black hole. When the mass approaches the correspondence mass \eqref{mcorr}, the temperature approaches the Hagedorn temperature, and the GR analysis breaks down. However, presumably the solution still exists, and one can ask where it lies in the mass vs temperature plot. For $d=6$ we proposed that it is given by the dashed blue line in figure \ref{d6thermo}, and it is interesting to ask what's the analog for $d<6$. 

Similarly, the HP solution corresponds to a line of solutions labeled by the temperature, or equivalently $m_\infty$ \eqref{mR}. When $m_\infty$ approaches $m_s$, the EFT analysis breaks down, but presumably the solution still exists, and one can ask where it lies in the mass vs temperature plot.  

In figure \ref{dneq6thermo} we propose one scenario for how these solutions can be continued. This proposal is motivated by the expectation that the structure of the solutions depends continuously on the dimension $d$. Thus, we start from the structure we proposed in section \ref{sec:md6} and ask how it is deformed when we go to $d$ slightly smaller than six. We already know from section \ref{sec:d=6}, that when we do this, the red curve in figure \ref{d6thermo} is deformed to the red curve in figure \ref{4ldl6}. 

It is thus natural, that it connects to the blue curve describing a small EBH, as depicted in the figure. If this does not happen, the red curve must continue to the small $M$ region, and there does not seem to be a natural solution for it to connect to there. The minimum of the $M(T)$ curve in figure \ref{4ldl6} is in this scenario dimension dependent, and approaches the origin as $d\to 6$.

Of course, the above argument is not conclusive, since it relies on assumptions about properties of the various solutions, and continuity in $d$. It would be interesting to establish these properties more conclusively. 

As the dimension $d$ decreases, we go in this scenario from the structure in figure \ref{4ldl6} for $4<d<6$, to that in figure \ref{d4} for $d=4$, and finally to that in figure \ref{dl4} for $d<4$. 

Note also that this scenario is in tension with the recent work \cite{Chen:2021dsw}, where it was argued that a continuous transition between the EBH and HP solutions is impossible in classical type II string theory. It would be interesting to understand this issue better.

So far, we focused on the case $d<6$. It is natural to ask what happens for $d>6$. The only solution that is known to exist in this case is the EBH, denoted by the solid blue line in the analog of figure \ref{d6thermo}. It is natural to conjecture that in the scaling limit $g_s\to 0$, with $M/m_s$ held fixed, this solution approaches the Euclidean two dimensional black hole, as for $d=6$. 
\begin{figure}[t]
\centering
 \includegraphics[width=0.5\textwidth]{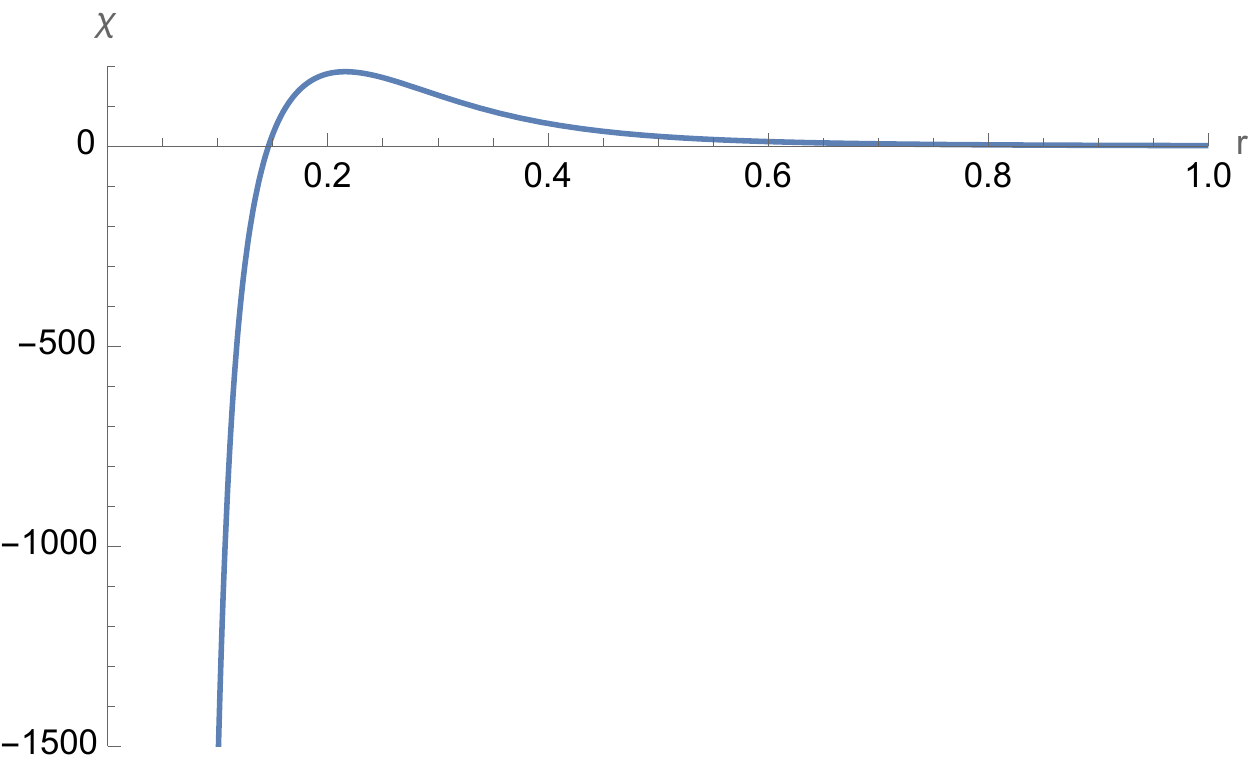}
 \caption{The solution $\chi(r)$ of \eqref{neweom} at $d=7$, with $C_\chi=1$ and $\sqrt{2}\alpha'/\kappa=1$.} 
\label{chid7}
\end{figure}

The main outstanding question is what happens to the red line in figure \ref{d6thermo} when we slightly increase the dimension $d$. Continuity in $d$ suggests that a solution with roughly this structure should still exist, however the EFT analysis does not yield such a solution. For example, if we assume that this solution has $R=R_H$, as in figure \ref{d6thermo}, and thus satisfies equations \eqref{chieqphi}, \eqref{neweom}, we find that for large $M$ the solution takes the form in figure \ref{chid7}. Thus, the EFT analysis breaks down for small $r$. 

Of course, a solution may still exist but require the inclusion of higher dimension operators in the EFT. This is natural from the point of view of the EFT \eqref{Sphichi}, since the $\varphi|\chi|^2$ coupling becomes irrelevant in this case, but we don't know how to go beyond \eqref{Sphichi} in a controlled manner. 

Another possibility is that the red line in figure \ref{d6thermo} disappears when $d$ goes above six. The advantage of this scenario is that one does not need to postulate the existence of additional solutions. The disadvantage is the discontinuity in $d$. We will leave the resolution of this issue to future work.

\vskip 1cm

\section*{Acknowledgements}

We thank O. Aharony, A. Giveon and E. Urbach for useful discussions and comments on the manuscript.
The work of BB and DK is supported in part by DOE grant DE-SC0009924. The work of DK is also supported in part by BSF grant number 2018068. DK thanks the Weizmann Institute for hospitality during part of the work on this paper.

%%%%%%%%%%%%%%%%%%%%%%%%%%%%%%%%%%%%%%
%%%%%%%%%%%%%%%%%%%%%%%%%%%%%%%%%%%%%%

%%%%%%%%%%%%%%%%%%%%%%%%%%%%%%%%%%%%%%
%%%%%%%%%%%%%%%%%%%%%%%%%%%%%%%%%%%%%%

\appendix

\section{Scaling analysis of the effective action}
\label{zetad6}

In this Appendix, we discuss the relative contribution of the terms in the effective action \eqref{Sphichi}, when evaluated on a solution to the equations of motion \eqref{eom0}, as in \cite{Chen:2021dsw}.

First, let $\chi_*(x),\varphi_*(x)$ be normalizable solutions of the equations of motion \eqref{eom0}. This is the case for example for the HP solution discussed in section \ref{sec:hpsol}. We are interested in exploring general field configurations that are related to $\chi_*(x),\varphi_*(x)$ by a simple rescaling. In particular, we consider the field configurations 
\begin{equation}
\label{chiphiscaling}
    \begin{split}
        &\chi(x)=\lambda \chi_*\left(\frac{x}{\gamma}\right),\\
        &\varphi(x)=\lambda^2\gamma^2 \varphi_*\left(\frac{x}{\gamma}\right).
    \end{split}
\end{equation}
The expression for $\varphi(x)$ in \eqref{chiphiscaling} is obtained by demanding that the scaling of $\varphi(x)$ is consistent with that of $\chi(x)$, as follows from the second equation in \eqref{eom0}. Under this rescaling, the action \eqref{Sphichi} becomes
	\begin{equation}
	\label{Sphichirescaled}
	\begin{split}
 	I_d(\lambda,\gamma)=\frac{\beta}{16\pi G_N}\int d^dx\left[\lambda^4\gamma^{d+2}(\nabla\varphi_*)^2+\lambda^2\gamma^{d-2}|\nabla \chi_*|^2+(\lambda^2\gamma^{d}m_{\infty}^2+\lambda^4\gamma^{d+2}\frac{\kappa}{\alpha'}\varphi_*)|\chi_*|^2\right]\ .
	\end{split}
	\end{equation}
Since $\chi_*(x),\varphi_*(x)$ is a saddle of the action, it must be that 
\begin{equation}
    \begin{split}
        \left.\frac{\partial I_d}{\partial \lambda}\right|_{\lambda=\gamma=1}=\left.\frac{\partial I_d}{\partial \gamma}\right|_{\lambda=\gamma=1}=0~,
    \end{split}
\end{equation}
from which we find that
\begin{equation}
\label{d6analysis}
    \begin{split}
        \int d^d x |\nabla \chi_*|^2=\frac{d-2}{6-d}m_{\infty}^2\int d^d x  \chi_*^2~.
    \end{split}
\end{equation}

Using the rescaling \eqref{rescalzeta} in \eqref{d6analysis}, we obtain
\begin{equation}
\label{d6analysisrescaled}
    \begin{split}
        \int d^d \tilde{x} |\tilde{\nabla} \tilde{\chi}|^2=\frac{d-2}{6-d}\zeta\int d^d \tilde{x}  \tilde{\chi}^2~,
    \end{split}
\end{equation}
where for simplicity we dropped the subscript, with the understanding that this equation is only valid for a normalizable solution of \eqref{eom0}.

Integrating by parts the l.h.s. of \eqref{d6analysisrescaled} and using the first equation in \eqref{dezeta}, we finally arrive at
\begin{equation}
    \begin{split}
        -\int d^d \tilde{x} \tilde{\chi}^2\left(\zeta+\tilde\varphi\right)=\frac{d-2}{6-d}\zeta \int d^d \tilde{x} \tilde{\chi}^2~,
    \end{split}
\end{equation}
where we dropped the boundary term since it vanishes for normalizable solutions. Rearranging this equation, we obtain \eqref{par}.

%%%%%%%%%%%%%%%%%%%%%%%%%%%%%%%%%%%%%%
%%%%%%%%%%%%%%%%%%%%%%%%%%%%%%%%%%%%%%

\vskip 2cm

\bibliographystyle{JHEP}
\bibliography{HP}
%%%%%%%%%%%%%%%%%%%%%%%%%%%%%%%%%%%%%%
%%%%%%%%%%%%%%%%%%%%%%%%%%%%%%%%%%%%%%

\end{document}